\UseRawInputEncoding
\let\labelindent\relax
\documentclass[letterpaper, 10 pt, conference]{ieeeconf}
\usepackage{tikz}
\usetikzlibrary{shapes.geometric, arrows, positioning, shadows, fit}

\IEEEoverridecommandlockouts

\usepackage{graphicx}
\usepackage{subfigure}
\usepackage{float}

\usepackage{subcaption}
\usepackage{amsmath}
\usepackage{amsfonts}
\usepackage{amsmath,amssymb}
\usepackage{color}

\usepackage{pifont}

\usepackage{makecell}
\usepackage{paralist}
\usepackage{diagbox}
\usepackage{seqsplit}
\usepackage{longtable}

\usepackage{algorithm}
\usepackage{algpseudocode}

\usepackage{booktabs} 
\usepackage{array}    

\usepackage{collcell}
\usepackage{colortbl}
\usepackage{multirow}
\usepackage{xcolor}
\usepackage{listings}
\usepackage{hyperref}
\hypersetup{hypertex=true,
	colorlinks=true,
	linkcolor=blue,
	anchorcolor=blue,
	citecolor=blue}

\title{$Orangutan$: A Multiscale Brain Emulation-Based Artificial Intelligence Framework for Dynamic Environments}
\author{Yong Xie\\
	Alibaba Group\\
	Hangzhou, China\\
	{\tt\small laola.xy@taobao.com}
}

\begin{document}
	
	\maketitle
	\thispagestyle{empty}
	\pagestyle{empty}
	
	\begin{abstract}
		
		Achieving General Artificial Intelligence (AGI) has long been a grand challenge in the field of AI, and brain-inspired computing is widely acknowledged as one of the most promising approaches to realize this goal. This paper introduces a novel brain-inspired AI framework, $Orangutan$. It simulates the structure and computational mechanisms of biological brains on multiple scales, encompassing multi-compartment neuron architectures, diverse synaptic connection modalities, neural microcircuits, cortical columns, and brain regions, as well as biochemical processes including facilitation, feedforward inhibition, short-term potentiation, and short-term depression, all grounded in solid neuroscience. Building upon these highly integrated brain-like mechanisms, I have developed a sensorimotor model that simulates human saccadic eye movements during object observation. The model's algorithmic efficacy was validated through testing with the observation of handwritten digit images.
		
	\end{abstract}
	
	\section{Introduction}
	
	As the ultimate goal in the field of artificial intelligence, General Artificial Intelligence (AGI) refers to AI that reaches human-level intelligence across all aspects. To achieve this ambition, the development of AI has drawn immense inspiration and insight from natural intelligence \cite{PMID:28728020}. Among these, AI frameworks represented by Deep Neural Networks (DNN) have achieved significant success in intelligent tasks such as image \cite{Krizhevsky2012ImageNetCW,Szegedy2015RethinkingTI} and speech recognition \cite{Hinton2012DeepNN}.
	
	Although DNNs have reached or even surpassed human performance in certain domains, they suffer from poor generalization, high dependency on data, lack of interpretability, and limited adaptability to environmental changes. These shortcomings pose significant challenges in advancing towards AGI. Many scholars attribute these limitations to the stark differences between the internal representations of DNNs and the neural activity patterns in biological brains, arguing that DNNs lack biological plausibility and struggle to match the high efficiency and dynamism of biological neural systems in information processing.
	
	In pursuit of AI pathways that more closely mimic brain operational mechanisms, numerous intelligence models and frameworks inspired by neuroscience have been proposed over the past decades. At the microscale, there are models like the Hodgkin–Huxley (H-H) model \cite{HODGKIN199025}, which simulates neuronal action potentials through nonlinear differential equations, and the simpler Integrate-and-Fire (IF) and Leaky Integrate-and-Fire (LIF) spiking neuron models. At the mesoscale, models like the Hierarchical Temporal Memory (HTM) \cite{Hawkins2004OnI} simulate brain cortical column structures, alongside theories such as the Thousand Brains Theory \cite{Vyshedskiy2022HawkinsJ2}. At the macroscale, spiking neural networks \cite{Maass1996NetworksOS} and the STDP learning algorithm \cite{Diehl2015UnsupervisedLO}, based on spiking neurons, have been developed. On the cognitive level, architectures such as Spaun \cite{Stewart2012SpaunAP}, ACT-R \cite{Anderson1998TheAC,Anderson1993RulesOT}, Leabra \cite{OReilly2012TheLC}, SAL \cite{Jilk2008SALAE}, Nengo \cite{Bekolay2014NengoAP}, NEST \cite{Gewaltig2007NESTS}, and BrainCog \cite{Zeng2022BrainCogAS} have been introduced.
	
	Despite each framework modeling different mechanisms at various scales of the brain, a vast range of computational mechanisms within the brain remains unexplored and insufficiently simulated. On the other hand, a unified architecture that effectively integrates brain-like mechanisms across multiple scales, similar to the brain, has yet to emerge. This paper posits that these two points are crucial for continued breakthroughs in AI and the progression towards AGI.
	
	To simulate as many brain mechanisms as possible across multiple scales, this paper introduces a new AI framework—$Orangutan$. At the microscale, it models the anatomical morphology of neurons, the transmission process of neural excitation, and various neurotransmitter actions; at the mesoscale, it simulates multiple types of neural synapses, neural microcircuits, and cortical column structures; at the macroscale, it models the structure of brain regions in the neocortex and the inter-regional projection relationships. Furthermore, based on its implemented model algorithms, it also simulates brain-like mechanisms such as population coding, winner-takes-all attention competition, and top-down feedback regulation. A detailed introduction will follow.
	
	\section{Framework Introduction}
	
\begin{figure*}
	\centering
	\includegraphics[width=1\linewidth]{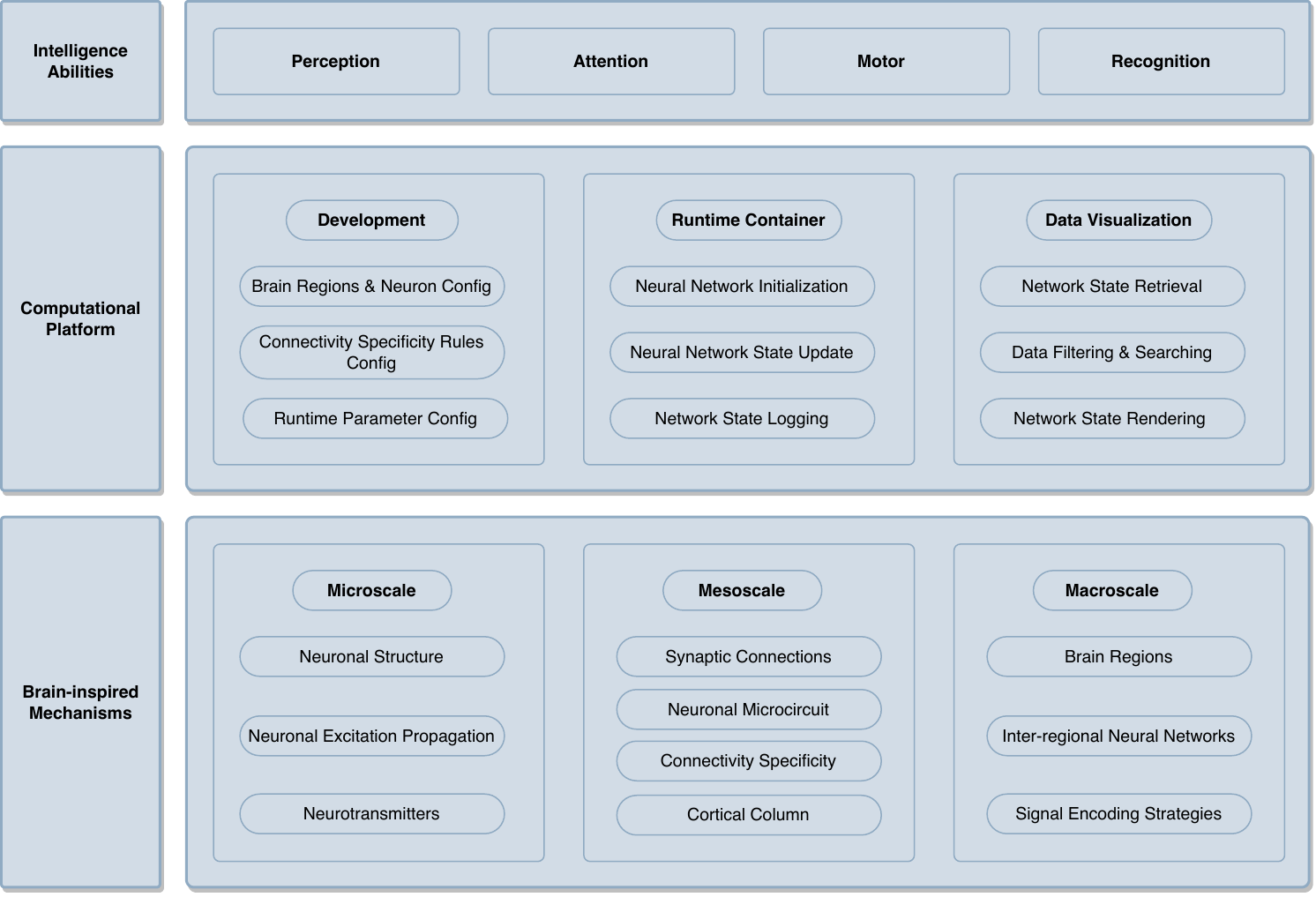}
	\caption{$Orangutan$ Architecture}
	\label{fig:framework}
\end{figure*}

	The Orangutan is a large primate species that inhabits the tropical rainforests of Asia. Not only are they remarkably intelligent, but they also possess a gentle disposition. These traits align closely with the long-term goal of this study: to achieve an artificial intelligence that is nearly on par with human intelligence and is friendly towards humans through high-fidelity simulation of the biological brain. The framework is thus named after this creature. To achieve this goal, $Orangutan$, utilizing the Python programming language, simulates the core mechanisms of the brain across the micro, meso, and macro scales. It aims to interpret their functions within the brain from an engineering and algorithmic perspective, with Figure \ref{fig:framework} depicting the main aspects of this endeavor. Additionally, considering the complexity involved in algorithm implementation, $Orangutan$ makes deliberate choices and adjustments in some biological details to strike a balance between biological plausibility and algorithmic feasibility. The mechanisms at each of these scales will be introduced subsequently.
	
	\subsection{Microscale Brain Simulation}
	
	At the microscale, the focus is on the most fundamental neuroscientific building blocks, including individual neurons, synapses, and the molecular and cellular processes within neurons.
	
\begin{figure}[!t]
	\centering
	\subfigure[] {\includegraphics[height=1.7in]{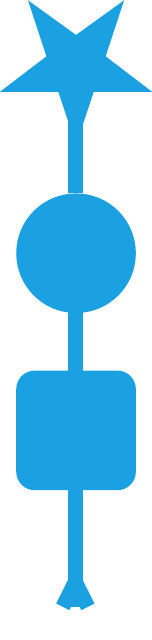}
		\label{fig:neuron:neuron}}
	\hfil
	\subfigure[] {\includegraphics[height=1.7in]{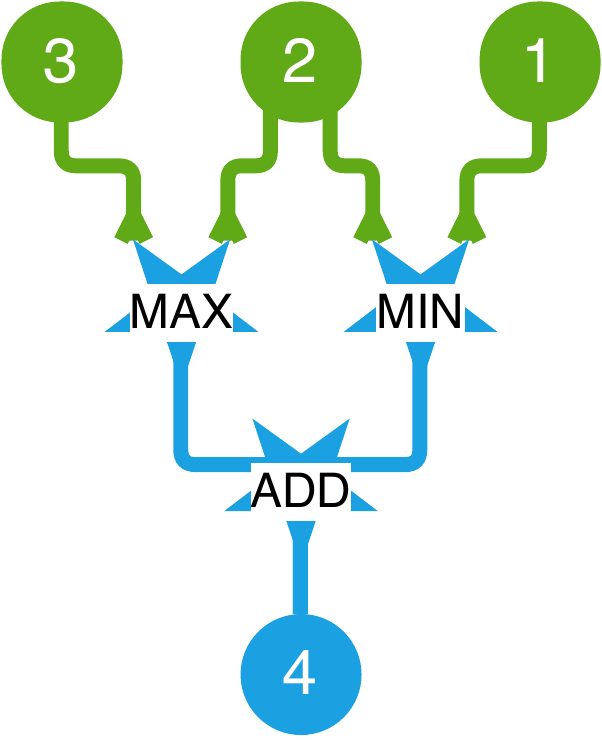} \label{fig:neuron:dendrite}}	
	\caption{Figure \subref{fig:neuron:neuron} depicts a typical $Orangutan$ multi-compartment neuron model. All compartments, from top to bottom, are sequentially arranged as dendrites (star), soma (round), axons (square), and synapses (the bulb-ended line). Figure \subref{fig:neuron:dendrite} shows a multi-compartment dendritic structure. Two primary dendritic compartments (MAX and MIN) simultaneously receive excitatory inputs from three presynaptic neurons (green), performing maximum and minimum calculations respectively, before passing their outputs to a secondary dendritic compartment (ADD) for summation. The numbers illustrate the excitatory inputs from the presynaptic neurons and the final excitation obtained by the soma (blue circle) after dendritic integration. }
	\label{fig:neuron}
\end{figure}

	Neurons serve as the basic computational units of neural networks. To more closely resemble the anatomical structure of biological neurons, $Orangutan$ has developed a multi-compartment neuron model \cite{Rall1962TheoryOP}, as opposed to the commonly used ``point'' neuron. A ``compartment'' refers to an abstract representation of a portion of a neuron, encompassing dendrites, soma (cell body), axons, synapses, etc., with Figure \ref{fig:neuron:neuron} illustrating the basic structure of an $Orangutan$ neuron. The neurocompartmental structure in $Orangutan$ offers considerable flexibility; in some simple computation scenarios, it allows developers to omit unnecessary dendrites or axons. When a neuron consists of only a single soma compartment, it becomes a typical ``point'' neuron. This simplification reflects $Orangutan$’s effort and attempt to balance biological plausibility with algorithmic feasibility by increasing algorithmic efficiency at the cost of a certain degree of simulation accuracy, while ensuring the computational outcomes of the neural network remain unchanged.
	
	$Orangutan$ simulates the electrophysiological and chemical properties of each compartment through a distinct set of attributes, providing developers the flexibility to adjust these values as shown in Table \ref{table:comp_props}. Leveraging this attribute set, $Orangutan$ models the dynamic process of excitation transmission between compartments. Initially, a compartment at its Resting Potential (RP) will generate a neural spike after receiving sufficient excitation. The frequency of the spike is calculated using Equation \ref{eq:frequency}, where $E$ denotes the current excitation of the compartment. The neural spike is transmitted progressively along the dendrites, soma, axons, and synapses. Equation \ref{eq:excite} illustrates the computation process for the excitation transmission across neuronal compartments, where ${E}'$ represents the excitation conveyed to the downstream compartment. After transmitting excitation, the compartment returns to the resting state.
	
	\begin{table*}[ht]
	\centering
	\caption{Neuronal compartment properties. All properties are represented as int or float for computational efficiency. Abbreviations are used for compartment names: A=Axon, S=Synapse, D=Dendrite.}
	\label{table:comp_props}
	\small
	\setlength{\tabcolsep}{4pt}
	\renewcommand{\arraystretch}{1.2}
	\begin{tabular}{
			p{2.3cm} 
			>{\centering\arraybackslash}p{1.1cm}
			>{\centering\arraybackslash}p{1cm} >{\centering\arraybackslash}p{1cm} >{\centering\arraybackslash}p{1.6cm} 
			p{9cm}
		}
		\hline
		Property & Symbol & Default Value & Exclu-sive & Synapses Inheritable & Description \\
		\hline
		\rowcolor{gray!50}
		\multicolumn{6}{l}{Fundamental Characteristics} \\
		type & N/A & N/A & All & \ding{55} & Compartment type, enumeration values found in Listing\ref{lst:vars_and_enums} under TYPE 
		\\
		region\_no & N/A & N/A & All & \ding{51} & Number of the brain region where the compartment is located 
		\\
		col\_no & N/A & N/A & All & \ding{51} & Number of the cortical column where the compartment is located 
		\\
		neuron\_no & N/A & N/A & All & \ding{55} & Number of the neuron where the compartment is located
		\\
		region\_row\_no & N/A & N/A & All & \ding{51} & Row number of the cortical column in the brain region 
		\\
		region\_hyper-\_col\_no & N/A & N/A & All & \ding{51} & Column number of the cortical column in the brain region 
		\\
		hyper\_col\_ind & N/A & N/A & All & \ding{51} & Index of the cortical hypercolumn within the brain region. \\
		mini\_col\_ind & N/A & N/A & All & \ding{51} & Index of the cortical minicolumn within the hypercolumn \\
		ind & N/A & N/A & All & \ding{55} & Index of the compartment in the entire neural network \\
		pre\_ind & N/A & N/A & All & \ding{55} & Index of the upstream compartment in the entire neural network \\
		post\_ind & N/A & N/A & All & \ding{55} & Index of the downstream compartment in the entire neural network \\
		\rowcolor{gray!50}
		\hline
		\multicolumn{6}{l}{Electrophysiological Characteristics} \\
		RP & RP & -65 & All & \ding{51} & Resting potential 
		\\
		excite & E & -65 & All & \ding{55} & Current excitation 
		\\
		tick\_spike-\_times & Freq & 0 & All & \ding{55} & Spike frequency generated by the current excitation \\
		step\_length & N/A & 1 & $A$ & \ding{55} & Axonal step length, influencing the timing of excitation reaching its target \\
		all\_or\_none & N/A & 0 & $S$ & \ding{51} & The synaptic all-or-none response pattern is modulated by both the stimulus of synaptic excitation, denoted as \(E'\), and the post-synaptic chamber excitation, denoted as E. This can be categorized as follows: 0 represents a non-binary response; 1 signifies a weak conditional binary response, transmitting the full excitation only when the synapse meets the condition \(|E'| > |E|\); 2 indicates a strong conditional binary response, whereby full excitation is transmitted when the synapse satisfies the condition \(|E'| \ge |E|\).
		\\
		\hline
		\rowcolor{gray!50}
		\multicolumn{6}{l}{Chemical Characteristics} \\
		Fa & Fa & 1 & All & \ding{55} & Facilitation effect experienced 
		\\
		Sh & Sh & 1 & All & \ding{55} & Shunting inhibition experienced 
		\\
		STP & STP & 1 & All & \ding{55} & Short-term potentiation experienced 
		\\
		STD & STD & 1 & All & \ding{55} & Short-term depression experienced 
		\\
		LTP & LTP & 1 & $S$ & \ding{55} & Long-term potentiation experienced 
		\\
		LTD & LTD & 1 & $S$ & \ding{55} & Long-term depression experienced
		\\
		transmitter-\_release\_sum& TRS & 65 & All & \ding{51} & Number of neurotransmitters released per spike 
		\\
		release\_type & N/A & N/A & $S$ & \ding{51} & Type of neurotransmitter released, with enumeration values available in Listing\ref{lst:vars_and_enums} under RELEASE\_TYPE. For both excitatory and inhibitory neurons, release\_type is ``excite'', differentiated by post\_sign=1/-1
		\\
		post\_sign & N/A & 1 & $S$/$D$ & \ding{51} & Distinguishes between excitatory and inhibitory neurons when release\_type is ``excite'', using 1 and -1.
		\\
		\hline
		\rowcolor{gray!50}
		\multicolumn{6}{l}{Neuroplasticity} \\
		marker & N/A & N/A & All & \ding{55} & Identifier number of neural cell adhesion molecules used to mark and recognize neuron compartments of the same type within a brain region \\
		produce\_marker-\_per\_spike & N/A & 0 & All & \ding{55} & Number of neural cell adhesion molecules produced per spike \\
		marker\_remain & N/A & 0 & All & \ding{55} & Current quantity of neural cell adhesion molecules, influenced by mechanisms such as spike generation, synaptic connection consumption, and tick-by-tick decay \\
		\hline
	\end{tabular}
\end{table*}
	
	\begin{equation}
		Freq=\max(\left\lfloor \frac{E-RP}{|RP|} \right\rfloor,0) 
		\label{eq:frequency}
	\end{equation}
	
	\begin{equation}
		{E}'=Freq\cfrac{\ max(Fa,STP)LTP} {\ max(Sh,STD)LTD}TRS
		\label{eq:excite}
	\end{equation}
	
	In $Orangutan$, the computational cycle during which excitation starts from a neuron's dendrites, moves through various compartments, and ultimately transfers to a compartment in another neuron, is defined as a ``tick''. Simulating the dynamic process of excitation transmission throughout the entire neural network is achieved by updating the state of each neuronal compartment tick by tick. However, compartments within the neural network are not merely passing excitation straightforwardly from upstream to downstream; instead, each type of compartment performs unique computational roles.
	
	Neuronal dendrites are made up of multiple dendritic compartments that can form branching and cascading structures \cite{Rall1959BranchingDT}. In addition to the operation of summation, inspired by the logical operations performed by the proximal dendrites of pyramidal neurons \cite{Polsky2004ComputationalSI}, $Orangutan$ has also established dendritic compartments that can carry out operations to determine maximum and minimum values. Figure \ref{fig:neuron:dendrite} demonstrates a secondary dendritic structure. These features enable dendrites to perform complex integration of incoming excitations and relay the integrated results to the soma.
	
	The soma serves as the primary site for representing neural signal intensity, simultaneously receiving excitations from its own dendrites or synapses from other neurons. It conducts a simple summation of these excitations before forwarding them to the axon. Additionally, in a ``point'' neuron model, the soma can also directly establish synaptic relationships with other neurons, both upstream and downstream.
	
	Axons are accountable for the output of neurons. $Orangutan$ emulates the characteristic of axons in biological brains where different neurons possess axons of varying lengths ($step\_length$) to precisely control the timing of excitation reaching the postsynaptic neuron, thus finely tuning the computational process of the neural network. Specifically, once an axonal compartment receives excitation at tick $t$, it becomes active and sends out spike signals at tick $t+step\_length$. Neuroscientifically, a biological neuron can have only one axon. To optimize the efficiency of the neural network operation, $Orangutan$ permits a neuron to have multiple axons, each operating independently with the capability to define distinct attributes such as $step\_length$, $post\_sign$, etc. An axon can simultaneously establish synapses with multiple postsynaptic neurons, facilitating the passage of excitation through these synapses.
	
	Synapses are the main sites of neuronal communication. In the biological brain, synapses influence the postsynaptic neuron by releasing different types of neurotransmitters, which commonly exert excitatory or inhibitory effects. $Orangutan$, following the approach of classical artificial neuron models, simplifies these effects into mathematical addition and subtraction operations \cite{McCulloch1990ALC,Sanger1958ThePA}. However, the signal transmission process in the biological brain is much more intricate, involving a variety of chemical actions. For example, facilitation ($Fa$) and shunting inhibition ($Sh$) can respectively amplify and diminish the strength of signal transmission \cite{Zucker2002ShorttermSP, Abbott2005DriversAM}. $Orangutan$ approximates these biochemical interactions through multiplicative and divisive mathematical operations, introducing non-linear computational elements to the network's computation process. Furthermore, these effects are transient (typically lasting only 1 tick in $Orangutan$), lending themselves well to synchronous computational scenarios between synapses. By comparison, short-term potentiation (STP) and short-term depression (STD) contribute to the neural network's short-term plasticity, just like facilitation and shunting inhibition, they enhance or diminish signal transmission but have longer durations \cite{Zucker2002ShorttermSP} (persisting for more than 1 tick in $Orangutan$). Such mechanisms can be applied to scenarios that require asynchronous computation between neurons, such as temporal computations and planning. Moreover, the impact of long-term potentiation (LTP) and long-term depression (LTD) mechanisms lasts even longer and reflects the network's long-term plasticity. Synapses can be categorized into stable and plastic based on whether the values of LTP or LTD alterations are permitted, with the former employed for algorithms unaffected by learned experience, such as basic perceptual functions, and the latter mainly applied within $Orangutan$’s learning algorithms.
	
	It is worth noting that $Orangutan$ does not directly simulate the neurotransmitters and specific molecular mechanisms behind these processes but emulates their ultimate impact on the particular compartments involved. This abstraction of molecular processes keeps the algorithmic complexity of simulating the neural network computation within reasonable bounds. The type of synapse is defined by the attribute $release\_type$, and the synapse's influence is reflected in changes to the related parameters of the postsynaptic compartment: the excitation of the postsynaptic compartment, impacted by excitatory and inhibitory synapses, will be incremented or decremented based on the excitation value $E^{'}$ transmitted from the presynaptic compartment (excluding dendritic compartments performing MAX and MIN operations). Similarly, postsynaptic compartments affected by facilitating synapses, shunting inhibitory synapses, short-term potentiation (STP), and short-term depression (STD) synapses will also have their corresponding property values ($Fa$, $Sh$, $STP$, $STD$) incremented by the value of $E^{'}$.
	
	Based on the aforementioned mechanisms, researchers can fulfill diverse computational requirements when developing neural networks by designing neurons with varying anatomical structures and utilizing different types of neurotransmitters for transmission.
	
	\subsection{Mesoscale Brain Simulation}
	
	Mesoscale studies focus on networks and circuits between neurons, examining how multiple neurons connect and collaborate to produce specific functions.
	
\begin{figure}[!t]
	\centering
	\includegraphics[width=1\linewidth]{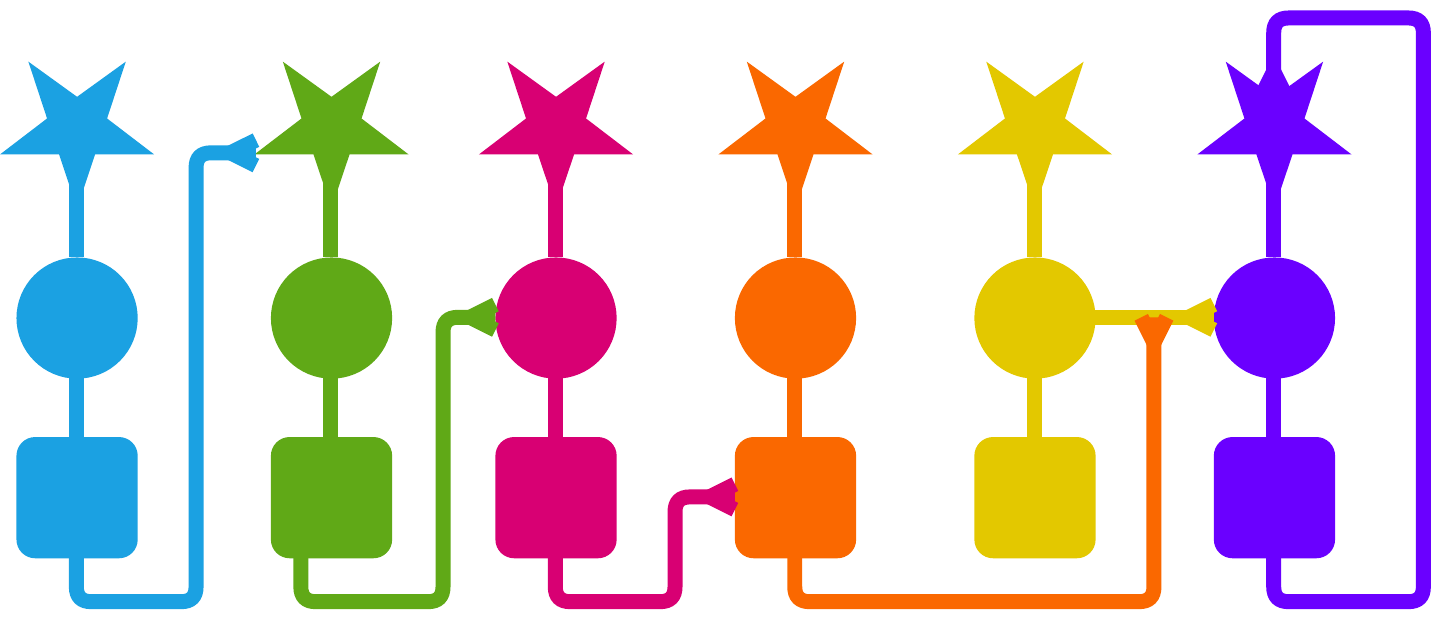}
	\caption{ Various synaptic connection types. From left to right, they are axo-dendritic, axo-somatic, axo-axonic, axo-synaptic, somato-somatic synapses, and autapses. To save space, other synaptic structures listed in Table \ref{table:synapse} are not shown in the figure. }
	\label{fig:synapse}
\end{figure}

	In the biological brain, there is an abundance of synaptic connection forms between neurons, including the common axo-dendritic synapses, as well as axo-somatic, axo-axonic, axo-synaptic, somato-somatic synapses, and autapses. $Orangutan$ meticulously simulates each of these types, as illustrated in Figure \ref{fig:synapse}. Beyond the most common axo-dendritic type, the other varieties of synapses in $Orangutan$’s neural networks serve different computational functions. Axo-somatic synapses enable the presynaptic neuron to circumvent the input signal integration mechanisms of the postsynaptic neuron's dendrites, directly affecting its soma. Axo-axonic synapses enable the presynaptic neuron to uniformly modulate the postsynaptic neuron's axon, thereby affecting the signal output of all synapses on that axon en masse. Axo-synaptic synapses, on the other hand, can finely regulate the output process of each synapse on the postsynaptic neuron. In addition to axonal synapses, the presynaptic neuron's soma may also directly establish synapses with any compartment of the postsynaptic neuron, such as somato-somatic synapses, functioning similarly to those formed by axons. This mechanism allows simple neuronal models that omit axonal compartments to still establish synapses with other neurons and participate in the neural network's computation process. Lastly, $Orangutan$ also simulates autaptic neurons found in the biological brain, where a neuron's axon or soma establishes synapses with its own compartments, which can provide a means for maintaining excitation or self-regulation. For a complete delineation of the synaptic relationships between compartments, refer to Table \ref{table:synapse}.
	
	\begin{lstlisting}[language=Python, caption=Main Variables and Enums, label=lst:vars_and_enums]
# Neuronal compartment types
TYPE = {
	soma: 1, 
	axon: 2,
	axon_end: 3,
	dendrite_min: 4,
	dendrite_max: 5,
	dendrite_add: 6,
}
# Types of excitatory actions of synaptic release
RELEASE_TYPE = {
	excite: 1, 
	Fa: 2, 
	Sh: 3, 
	STP: 4, 
	STD: 5,
}
# The Full Extent of the Global Receptive Field in Visual Models
VISUAL_FIELD_WH = (28, 28)
# Orientation enumeration values defined within a polar coordinate system where the upward direction corresponds to 0 or 360 degrees.
ORIENTS = [22.5, 45., 67.5, 90., 112.5, 135., 157.5, 180., 202.5, 
225., 247.5, 270., 292.5, 315., 337.5, 360.]
# The number of orientations.
ORIENT_SUM = 16
# The minimum spacing between orientations.
MIN_ORIENT = 22.5
# Scale enumeration values in pixels. 
RECEPTIVE_FIELD_LEVELS = [1, 3, 5, 7, 9, 11, 13, 15, 17, 19, 21]
# Angle enumeration values.
ANGLES = [22.5, 45., 67.5, 90., 112.5, 135., 157.5, 180.]
\end{lstlisting}
	
	\begin{table}[ht]
	\centering
	\caption{Permitted Synaptic Relationships in Orangutan}
	\label{table:synapse}
	\small
	\setlength{\tabcolsep}{2pt}
	\begin{tabular}{|c|c|c|c|c|}
		\hline
		\diagbox{\parbox{0.5cm}{Presynaptic}}{\parbox{1.8cm}{Postsynaptic}} & Dendrite & Soma & Axon & Synapse \\
		\hline
		\rowcolor{gray!50}
		\multicolumn{5}{|c|}{Heterosynapses} \\
		Axon & \ding{51} & \ding{51} & \ding{51} & \ding{51} \\
		\hline
		Soma & \ding{51} & \ding{51} & \ding{51} & \ding{51} \\
		\hline
		\rowcolor{gray!50}
		\multicolumn{5}{|c|}{Autapses} \\
		Axon  & \ding{51} & \ding{51} & \ding{55} & \ding{51} \\
		\hline
		Soma  & \ding{51} & \ding{55} & \ding{51} & \ding{51} \\
		\hline
	\end{tabular}
\end{table}

\begin{figure}[!t]
	\centering
	\subfigure[] {\includegraphics[height=1in]{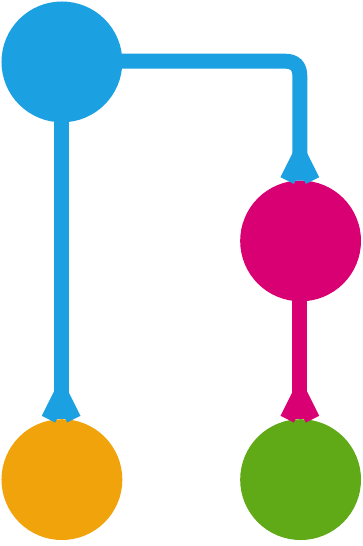}\label{fig:micro_circuit:side_inhibit}}
	\hfill
	\subfigure[] {\includegraphics[height=1in]{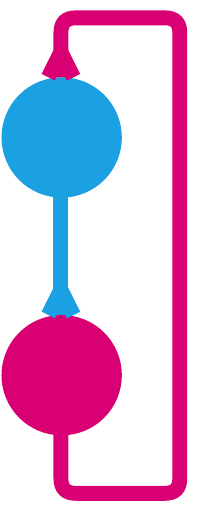}\label{fig:micro_circuit:back_inhibit}}
	\hfill
	\subfigure[] {\includegraphics[height=1in]{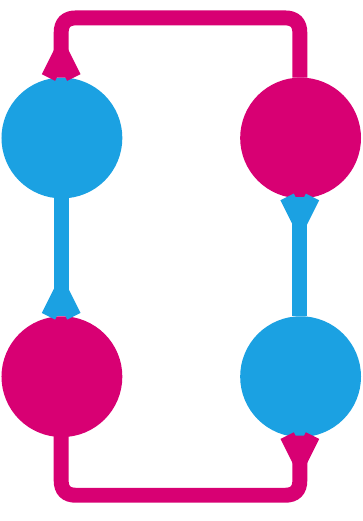}\label{fig:micro_circuit:mutual_inhibit}}
	\hfill
	\subfigure[] {\includegraphics[height=1in]{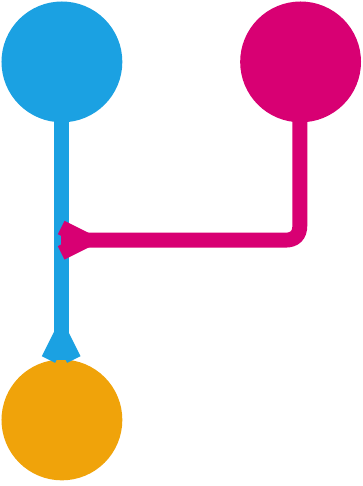}\label{fig:micro_circuit:not_logic}}
	\\
	\subfigure[] {\includegraphics[height=1in]{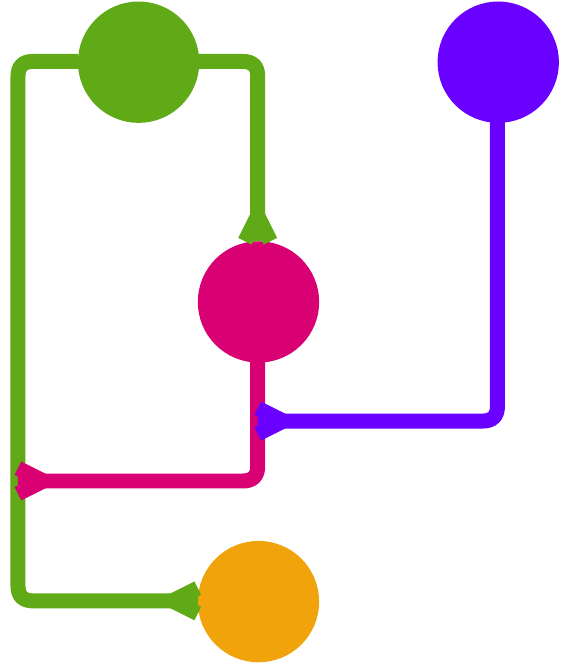}\label{fig:micro_circuit:and_logic}}
	\hfill
	\subfigure[] {\includegraphics[height=1in]{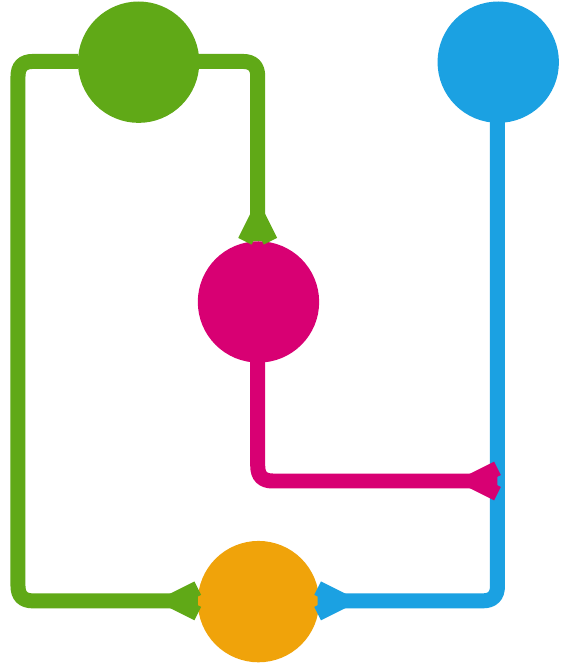}\label{fig:micro_circuit:or_logic}}
	\hfill
	\subfigure[] {\includegraphics[height=1in]{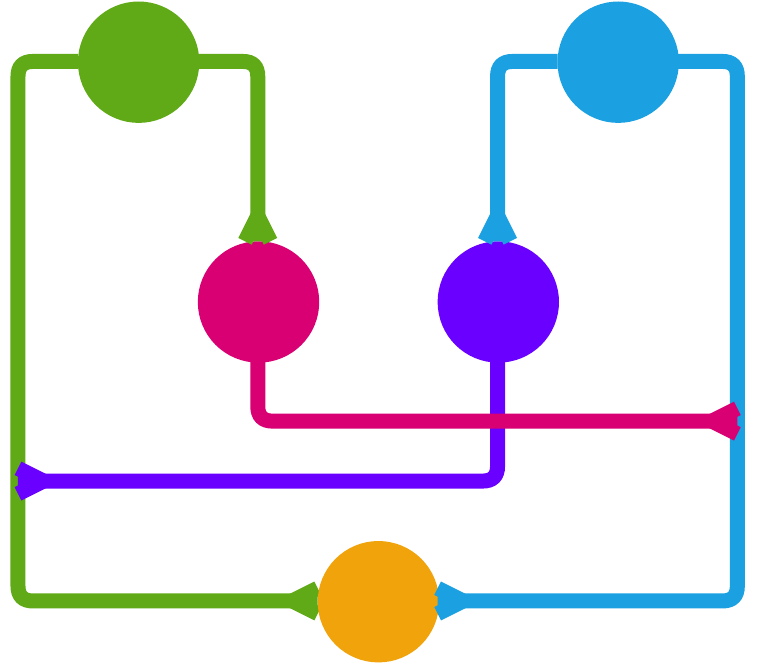}\label{fig:micro_circuit:xor_logic}}
	\caption{Neural Microcircuits, where both the red and purple neurons are inhibitory neurons. In practical scenarios, $Orangutan$ simplifies neural circuits by replacing excited inhibitory interneurons with the inhibitory axons of presynaptic neurons, thereby reducing the neural network's complexity and operational overhead. \subref{fig:micro_circuit:side_inhibit} Lateral inhibition features a neuron (blue) that excites a target (yellow), whilst inhibiting nearby cells (green) via an interneuron (red). \subref{fig:micro_circuit:back_inhibit} In recurrent inhibition, the neuron (blue) inhibits itself through an interneuron (red). \subref{fig:micro_circuit:mutual_inhibit} For mutual inhibition, each neuron (blue) suppresses the other via an interneuron (red). \subref{fig:micro_circuit:not_logic} A NOT gate arises when the excitatory neuron’s (blue) activity is prevented by an inhibitory neuron (red). \subref{fig:micro_circuit:and_logic} In an AND gate circuit, the green excitatory neuron activates both yellow and red neurons equally, but red's inhibition prevents yellow from activating. Only with both green and active inhibitory purple neurons does the yellow output neuron activate. \subref{fig:micro_circuit:or_logic} An OR gate triggers the output (yellow) when either the green or blue cell is active. When active together, the green neuron uses the red neuron to inhibit the blue synapse, protecting the yellow neuron from overactivation. \subref{fig:micro_circuit:xor_logic} An XOR gate activates the output (yellow) only when activity is present in one of two antagonistic neurons (green, blue).}
	\label{fig:micro_circuit}
\end{figure}

	Building upon the rich variety of synaptic connection types, $Orangutan$ further simulates numerous classic neural microcircuit structures found in the biological brain, as depicted in Figure \ref{fig:micro_circuit}. These microcircuits act as fundamental algorithmic units within larger-scale neural networks, serving various functions such as lateral inhibitory circuits for enhancing signal contrast \ref{fig:micro_circuit:side_inhibit}, recurrent inhibitory circuits for feedback regulation \ref{fig:micro_circuit:back_inhibit}, and mutual inhibitory circuits for establishing antagonistic actions or competitive relationships between neurons \ref{fig:micro_circuit:mutual_inhibit} \cite{Luo2021ArchitecturesON}. Among these, mutual inhibitory circuits implement a Winner-Takes-All (WTA) competitive mechanism, where at any given moment only one neuron in the circuit is active while the others are silenced due to inhibitory effects. $Orangutan$ extends the applicability of the WTA mechanism to include synaptic inhibition, such that in an inhibitory axo-synaptic synapse, if the excitation of the presynaptic neuron exceeds that of the postsynaptic neuron, the former can completely inhibit the latter's synapse, and vice versa. $Orangutan$ achieves this all-or-none discharge characteristic in inhibitory synaptic compartments through the attribute $all\_or\_none$. This mechanism allows the presynaptic neuron to either fully inhibit or not affect a specific synapse of the postsynaptic neuron, without influencing the soma of the latter, thus ensuring that other synapses of the postsynaptic neuron can still output signals normally. This holds significant utility in scenarios requiring such refined regulation. It is noteworthy that in cases where the excitation levels of neurons are equal, the inhibitory circuit may exist in a critical state. For the axo-synaptic synapse, this means the postsynaptic neuron could either not at all inhibit or fully inhibit the postsynaptic compartment, thus representing two independent critical states. $Orangutan$ distinguishes these by setting $all\_or\_none$ to 1 or 2, termed as weak all-or-none inhibition and strong all-or-none inhibition, respectively. These critical state all-or-none inhibition mechanisms, akin to highly sensitive triggers, require only minimal perturbations to disrupt the equilibrium and are heavily utilized in the brain-like models subsequently introduced.
	
	Beyond these typical microcircuits found in the biological brain, $Orangutan$ also simulates microcircuits capable of performing logical operations, including AND gates \ref{fig:micro_circuit:and_logic}, OR gates \ref{fig:micro_circuit:or_logic}, NOT gates \ref{fig:micro_circuit:not_logic}, and XOR gates \ref{fig:micro_circuit:xor_logic}. From the perspective of a single spike, whether a neuron fires at a certain moment corresponds to a binary 1 or 0, making these logical microcircuits equivalent to the logic gates found in traditional circuits. In reality, neurons within these circuits often exhibit different pulsing frequencies, and their released excitation is regulated by neurotransmitters. Therefore, from a broader time scale, a NOT gate microcircuit performs a subtraction operation, an AND gate microcircuit operates as a minimum value calculation, and an OR gate microcircuit as a maximum value calculation.
	
	The above neural microcircuits demonstrate the sparsification of connections between $Orangutan$ neurons, meaning each neuronal compartment is connected only to a specific set of compartments within the neural network. Unlike classical fully connected artificial neural networks, sparse networks not only help to enhance overall operational efficiency but also align more closely with our observations of biological neural networks \cite{Song2005HighlyNF}. To establish precise synaptic connections between specified neurons, $Orangutan$ adopts a connectivity specificity mechanism inspired by the biological brain \cite{Sperry1963CHEMOAFFINITYIT} and has developed a rule-based system: based on attributes such as $region\_no$, $neuron\_no$ and $col\_no$ (see Table \ref{table:comp_props}), developers can obtain each compartment's index value within the neural network and design rule functions that precisely describe the presynaptic and postsynaptic compartments to guide synapse formation. This rule system applies to both the genesis of stable and plastic synapses; stable synapses are created at the initialization phase of the neural network, whereas plastic synapses are dynamically generated during network operation. Attributes marked as synapse only in Table \ref{table:comp_props} are defined in the upstream compartment but do not act on it; instead, they are inherited by the nascent synapse upon its formation. Additionally, $Orangutan$ allows developers to assign values to the attributes of these new synapses within the rule functions. Networks established through these pre-encoded rules can perform specific algorithms and possess good interpretability.
	
	Based on the architecture of neural microcircuits and connectivity specificity mechanisms, $Orangutan$ further simulates the columnar structure of the cerebral neocortex. Cortical columns are the basic functional units of the neocortex, composed of several neurons with the same $col\_no$, exhibiting more complex information processing capabilities than individual neurons. Additionally, cortical columns in the biological brain often possess specialized functions; for instance, in the visual cortex, different columns may be sensitive to edges of different orientations or stimuli of varying colors \cite{Hubel1962ReceptiveFB,Hubel1968ReceptiveFA}. $Orangutan$ adheres to this discovery in practical applications of cortical columns. The computational function of cortical columns is realized by local neural networks formed within the column through the mechanism of connection specificity. These neurons can not only share input signals but also, after a series of computations, transmit the processed results to other cortical columns via intercolumnar synapses, generating more complex computational activities at a macroscopic scale.
	
	\subsection{Macroscale Brain Simulation}
	
	At the macroscopic scale, the focus is on the structure and function of the entire brain or the entire nervous system, including the connections between different brain regions (the brain connectome) and how these connections support complex cognitive functions and behaviors. At this scale, numerous cortical columns make up the brain region structures at the macroscopic level.
	
	Brain regions in $Orangutan$ are computational modules with a single function, designed to process specific signals or perform particular functions. Inspired by relevant research \cite{Mountcastle1978AnOP}, cortical columns within an $Orangutan$ brain region have similar intra-column network structures and algorithmic logic, whereas cortical column structures vary between different brain regions. Furthermore, cortical columns within the same brain region can establish cross-column synaptic connections, forming neural networks at the regional scale to perform complex module-level algorithms. Additionally, cortical columns across different brain regions can also establish inter-regional synaptic projections, linking multiple computational modules in series to form more complex information processing pathways and computational systems.
	
	The multitude of neurons within a brain region collectively establish a complex encoding mechanism for signal input, exhibiting characteristics similar to the sparse and population coding found in the brain. Sparse coding \cite{Olshausen2004SparseCO} refers to a neural system's strategy of representing information where, due to sparse connections between neurons, only a few neurons are activated while most remain silent. This encoding strategy is considered efficient as it reduces energy consumption and enhances the selectivity and robustness of representations. Sparse coding ensures that neuronal response patterns have a high degree of selective specificity, with a neuron responding strongly to only a few stimuli. Population coding \cite{Averbeck2006NeuralCP}, on the other hand, represents information through the overall activity patterns of neuron groups, rather than the activity of individual neurons. In population coding, a single neuron may respond to multiple inputs, and different combinations of neurons can encode more complex information. This strategy allows the brain to represent rich and detailed information through multi-dimensional activity patterns of neuronal groups, increasing the economic efficiency of the model's algorithms. For example, the joint activity of two adjacent neurons can theoretically represent any value between the two numerical values they encode, such as the combined activity of two neurons representing red and yellow to indicate the color orange.
	
	\subsection{Brain Simulation of Learning Mechanisms}
	
	Learning mechanisms are one of the core elements of artificial intelligence, and $Orangutan$'s learning mechanisms are still in an exploratory phase. Currently, there is a degree of simulation for various learning and memory-related mechanisms in the brain, including Hebbian learning and predictive coding. It is hoped that through the high-level integration of these mechanisms, a learning ability can ultimately be achieved that allows for online and local learning, possesses high sample efficiency and interpretability, and approaches human-level capabilities.
	
	$Orangutan$'s learning mechanism will follow Hebb's rule, which states that synapses are formed between simultaneously active neurons. In the biological brain, neuronal adhesion molecules are involved in regulating this synaptic plasticity, which is very important for learning and memory. $Orangutan$ simulates the type of neuronal adhesion molecules and their production and depletion dynamics using three attributes: $marker$, $produce\_marker\_per\_spike$, and $marker\_remain$. The $marker$ serves to label specific types of neural cells, $produce\_marker\_per\_spike$ defines the number of neuronal adhesion molecules released per spike, and $marker\_remain$ indicates the remaining abundance of these molecules, with more details available in Table \ref{table:comp_props}. Furthermore, through the rule system of $Orangutan$, developers can write a number of directional rules specifying the $marker$ of presynaptic and postsynaptic neurons that can dynamically establish and strengthen synapses. During the operation of the neural network, when neurons that meet the criteria are active in the same tick, they will establish new synapses or change the strength of existing ones. For the learning mechanism, the rule mechanism can constrain the scope of neural plasticity. By limiting which neurons can or cannot form connections with each other, potentially harmful cognitions or skills beyond what is desired can be avoided as the agent explores the world, contributing to safer and more controllable artificial intelligence.
	
\begin{figure}[t]
	\centering 
	\includegraphics[height=3cm]{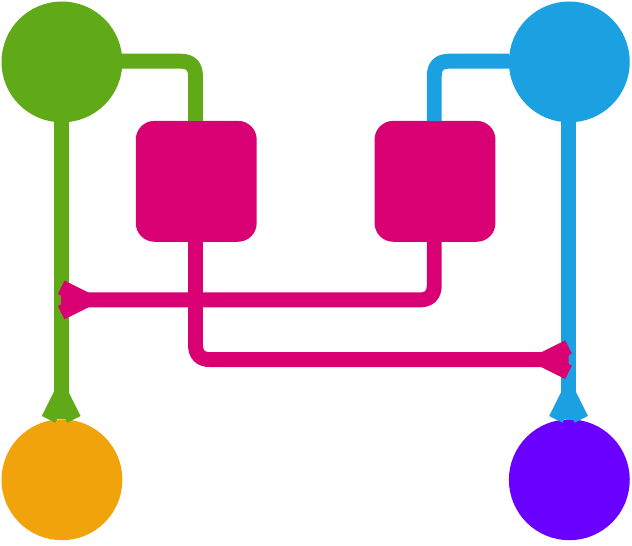}
	\caption{Predictive Coding Neural Circuit. Prediction cell (green) and input cell (blue) separately excite the negative prediction bias cell (orange) and positive prediction bias cell (purple), respectively. They form a NOT gate microcircuit with each other through their respective inhibitory axons and synapses (red). When the excitation of the prediction neuron is greater than that of the input neuron, the negative prediction bias neuron is activated, and vice versa for the positive prediction bias neuron. If the excitations of both are exactly equal, then neither of the prediction bias neurons will be activated.}
	\label{fig:predict_coding}
\end{figure}

	In addition to Hebbian learning, $Orangutan$ also draws from the theory of predictive coding in neuroscience to simulate the dynamic changes in synaptic Long-Term Potentiation ($LTP$) and Long-Term Depression ($LTD$). The $LTP$ and $LTD$ of plastic synapses are regulated by neurons with negative and positive prediction biases, respectively. The former becomes active and induces an increase in $LTP$ when outcomes exceed expectations, while the latter becomes active and leads to an increase in $LTD$ when outcomes are below expectations. These types of neurons are activated through specific neural circuits, as shown in Figure \ref{fig:predict_coding}.
	
	The specific implementation of the above learning mechanisms is subject to change as $Orangutan$ continues to explore the brain, and thus this article does not intend to reveal too much detail before the technology matures. Additionally, I am also attempting to introduce more brain-like learning mechanisms, such as hippocampal replay and dendritic spine resource competition.
	
	\section{Model Introduction}
	
\begin{figure}[t]
	\centering 
	\includegraphics[width=0.9\linewidth]{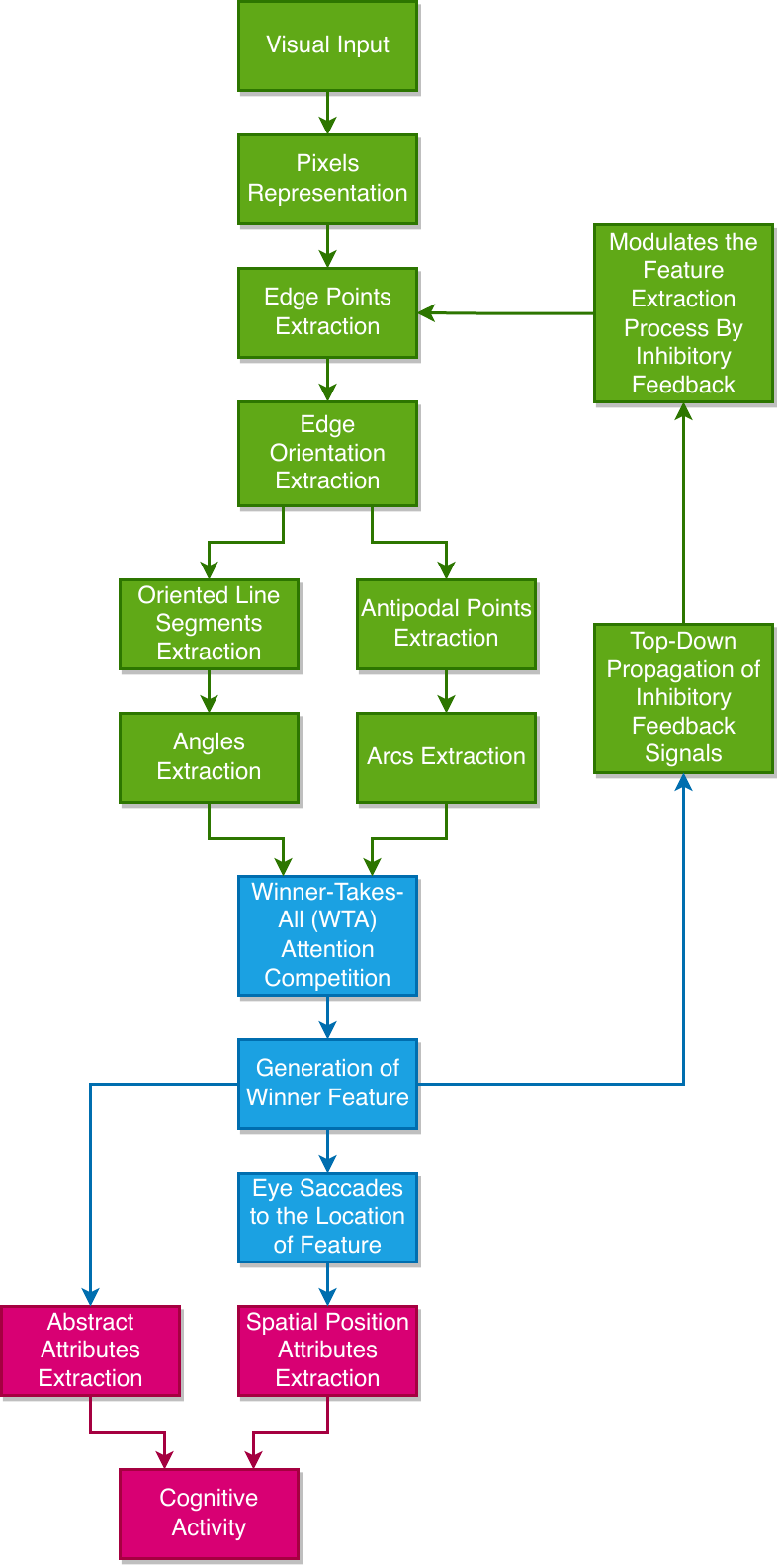}
	\caption{Model Workflow. It primarily involves three components: perception (green), motion (blue), and abstraction (red).}
	\label{fig:model_workflow}
\end{figure}

	To validate the algorithmic efficacy of $Orangutan$, I have developed a large neural network model that utilizes most of the mechanisms provided by $Orangutan$. It comprises 13 brain regions, approximately 3.7 million neurons, and about 56 million compartments. This model simulates the saccadic movements observed when the human eye views objects. At first glance, saccades seem to be a simple mechanical process, but according to Moravec's paradox (tasks that are easy for humans are hard for machines and vice versa), the complex neural mechanisms and dynamic processes involved present a significant challenge for any artificial intelligence system attempting to simulate similar effects. In classical visual theories of neuroscience, the biological brain processes visual signals primarily through the ``what'' and ``where'' pathways, with the ``what'' pathway responsible for processing pattern information, and the ``where'' pathway for processing spatial information and motion control \cite{Mishkin1983ObjectVA}. The operation of $Orangutan$ is broadly divided into perception, motion, and abstraction, as shown in Figure \ref{fig:model_workflow}. The perceptual part simulates the classical hierarchical feature extraction process of the ``what'' pathway \cite{Hubel1962ReceptiveFB}, while the motion part mimics the vision-guided motor-related processes of the ``where'' pathway. The perception phase provides potential feature targets for the motion phase, which in turn regulates the feature extraction process of the perception phase. These two phases influence each other and work in coordination, jointly realizing a neural network dynamic cycle of perception and motion. Meanwhile, accompanying the perception-motion process, the model continuously performs spatial perception of feature locations and extraction of abstract attributes of features, providing a more comprehensive representation of objects for downstream cognitive activities. Finally, it is important to note that since this model does not include learning mechanisms, all synapses in the neural network are stable synapses. Next, I will introduce the working principles of these three stages separately.
	
	I utilized the MNIST handwriting digits dataset as the signal input source for this model. As one of the most renowned and extensively used datasets in the field of artificial intelligence, MNIST has been employed to validate numerous AI technologies in their nascent stages of development, including Convolutional Neural Networks (CNNs) \cite{LeCun1998GradientbasedLA} and Dropout \cite{JMLR:v15:srivastava14a}, among others. The following sections will elucidate the working principles of the three stages of this model on MNIST samples.
	
	\subsection{Perception}
	
	The perceptual capability of this model aims to extract top-level visual features from images. For this model, two of the most common top-level features were defined: arcs and angles, which have good generalizability for the visual patterns of handwritten digits and can describe almost all Mnist image cases. Related research has also revealed the presence of neurons in the primary visual cortex that have high selectivity for these two types of features \cite{Tang2018ComplexPS}.
	
	The model has developed multiple brain regions to work together on hierarchical feature extraction. It is important to note that the definition of these brain regions does not follow the typical division found in the biological brain but is instead aligned with the actual algorithmic requirements and the features to be extracted—that is, each brain region is responsible for extracting only one type of feature.
	
	\subsubsection{Representing Pixel Points}
	
\begin{figure}[!t]
	\centering
	\subfigure[] {\includegraphics[width=0.42\linewidth]{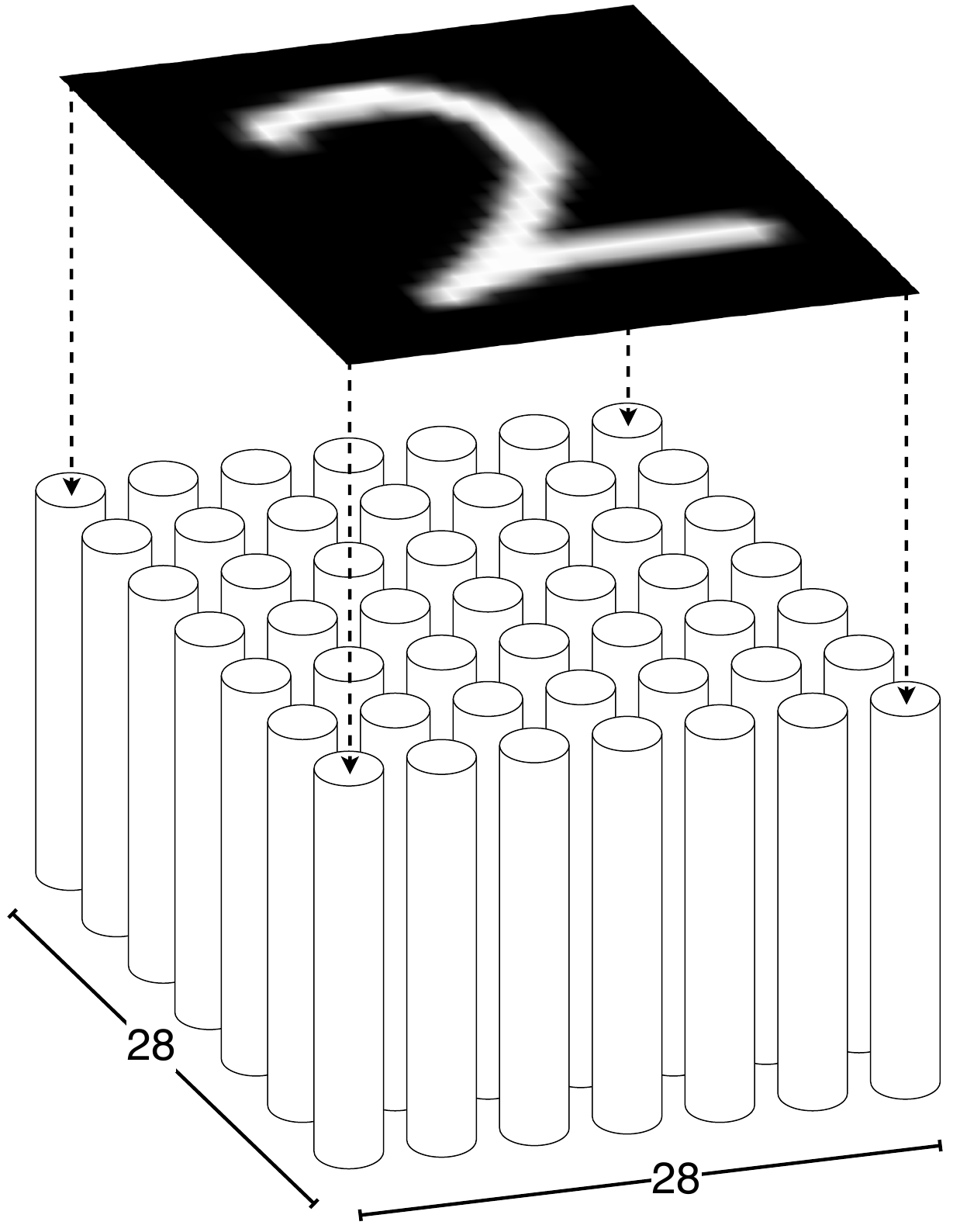} \label{fig:perception:topology}}
	\hfill
	\subfigure[] {\includegraphics[width=0.5\linewidth]{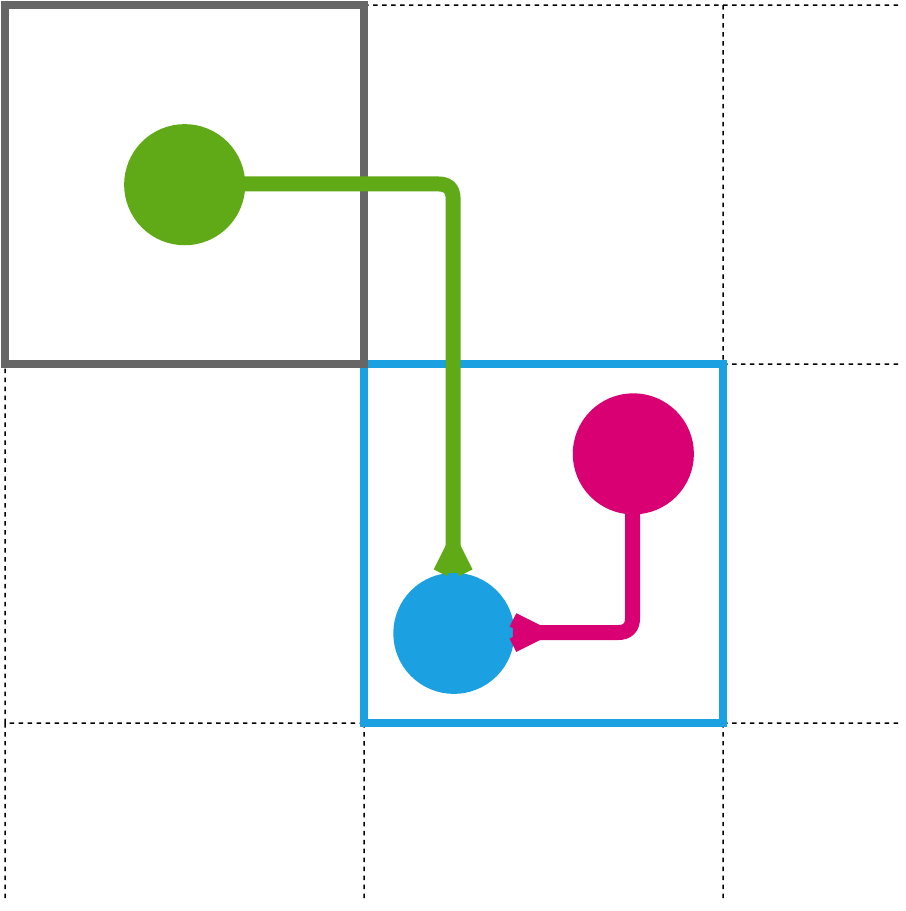} \label{fig:perception:edge_circuit}}
	\hfill
	\caption{
		\subref{fig:perception:topology} Each of the 28x28 hypercolumns in the brain region involved in feature extraction topologically maps to an image pixel, processing localized visual features. Dashed arrows highlight this correspondence at the image and cortical column peripheries.
		\subref{fig:perception:edge_circuit} Central-surround antagonistic neural circuit for edge point extraction. Within a 3x3 cortical column array, the $CEdg^{o=135}$ (blue) located in the central receptive field (outlined by the blue box) integrates excitatory inputs from $CPxl$ (green) in the peripheral receptive field (outlined by the black box) with inhibitory inputs from $CPxl$ (red) in the central receptive field. As the peripheral receptive field of $CEdg^{o=135}$ spans only a single grid space, the circuit includes only one excitatory neuron. However, for receptive fields that cover multiple grid spaces, such as the peripheral receptive field of the first $CEdg^{o=22.5}$ shown in Figure \ref{fig:receptive_field:edge_point}, there are multiple excitatory neurons present within the circuit, and the excitatory weights transmitted are directly proportional to the grayscale levels of the corresponding grid spaces.
	}
\end{figure}

	In the publicly available Mnist dataset, each image is a 28x28 grayscale picture, with each pixel's grayscale value ranging from 0 to 255. To represent these pixel points, this model has established a brain region for processing feature points, consisting of 28x28 cortical columns. For ease of description, $Orangutan$ utilizes a Cartesian coordinate system with the y-axis pointing downwards and the x-axis to the right to mark the position of each cortical column, placing the top-leftmost cortical column at the coordinate origin. The arrangement of cortical columns within the brain region forms a one-to-one topological mapping relationship with the pixels on the image. This is analogous to the topological projection relationship between retinal ganglion cells and the lateral geniculate nucleus of the thalamus \cite{Hubel1962ReceptiveFB,Hubel1968ReceptiveFA}, as shown in Figure \ref{fig:perception:topology}. Each point column contains a neuron representing the corresponding pixel point, termed a pixel cell, denoted as $CPxl$, with more details provided in Table \ref{table:regions}.
	
	\begin{figure}[!t]
	\centering
	\subfigure[] {\includegraphics[width=0.35\linewidth]{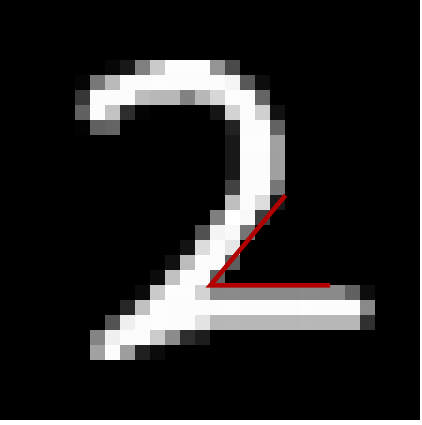} \label{fig:feature:ang}}
	\hfil
	\subfigure[] {\includegraphics[width=0.35\linewidth]{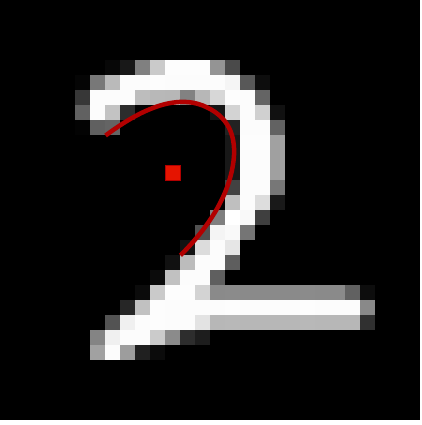} \label{fig:feature:arc}}
	\caption{Two types of high-level features defined by this model:
		\subref{fig:feature:ang} Feature angles.
		\subref{fig:feature:arc} Feature arcs. The red squares represent the centroids of the arcs.}
	\label{fig:feature}
\end{figure}
	
\begin{figure}
	\centering
	\subfigure[] {\includegraphics[height=1in]{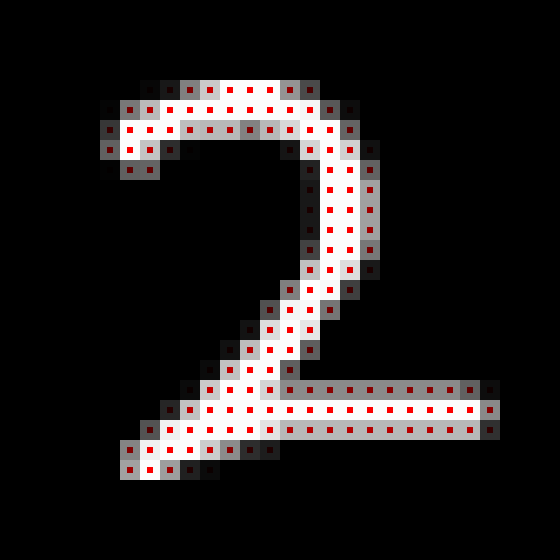} \label{fig:perception_res:pixel}}
	\hfill
	\subfigure[] {\includegraphics[height=1in]{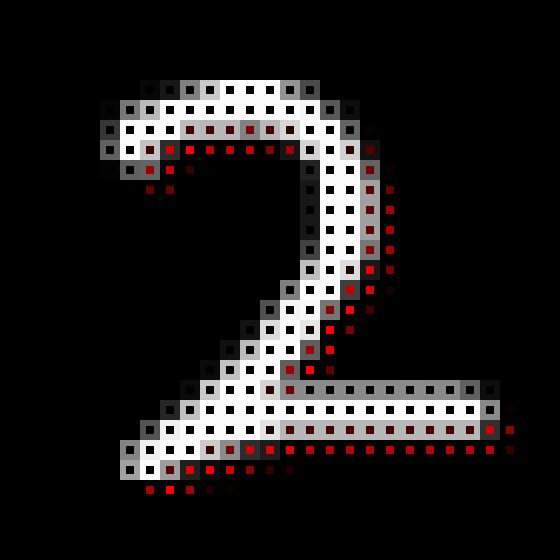} \label{fig:perception_res:edge_dot}}
	\hfill
	\subfigure[] {\includegraphics[height=1in]{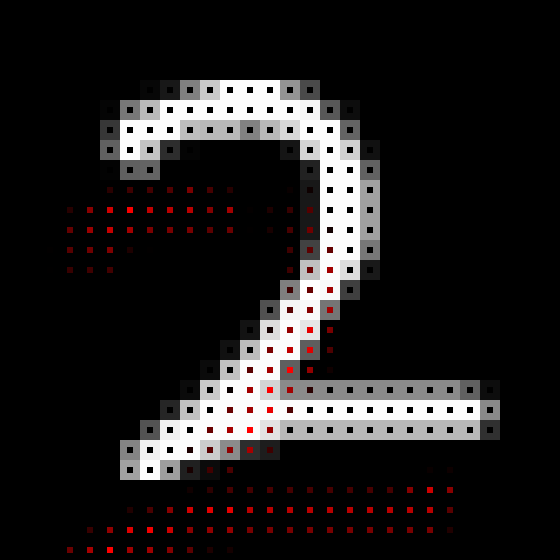} \label{fig:perception_res:orient}}
	\\
	\subfigure[] {\includegraphics[height=1in]{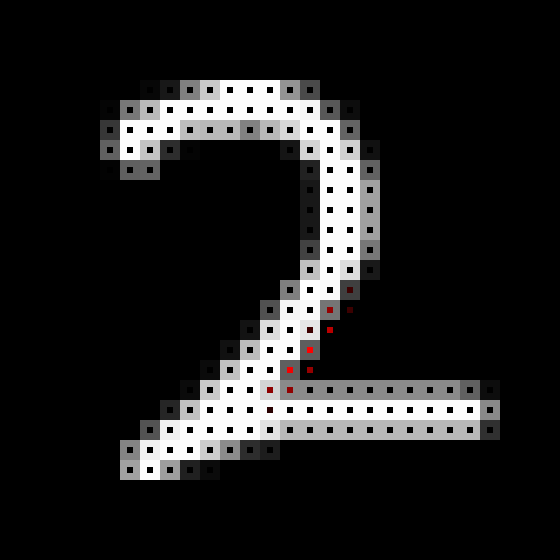} \label{fig:perception_res:seg}}
	\hfill
	\subfigure[] {\includegraphics[height=1in]{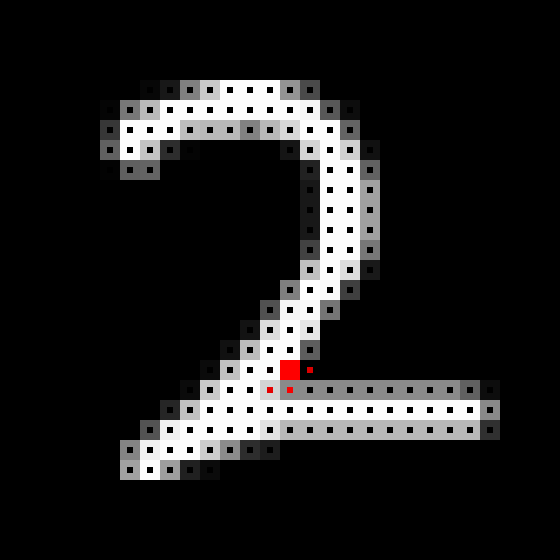} \label{fig:perception_res:ang}}
	\hfill
	\subfigure[] {\includegraphics[height=1in]{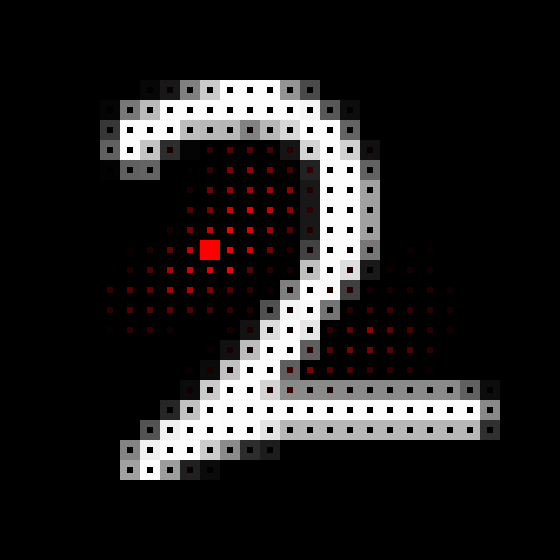} \label{fig:perception_res:arc}}
	\caption{Extraction results of features at various hierarchical levels on image 2\_72.
		\subref{fig:perception_res:pixel} Neurons (red spot) representing pixel points (white). The whiter the pixel, the brighter the red spot, indicating higher neuron excitation; conversely, the darker the spot, the lower the excitation.
		\subref{fig:perception_res:edge_dot} Activity of \(CEdg^{o=135}\)
		\subref{fig:perception_res:orient} Activity of \(COri^{s=9,o=45,a=90}\)
		\subref{fig:perception_res:seg} Activity of \(CSeg^{s=5,o=45,a=90}\)
		\subref{fig:perception_res:ang} Activity of \(CAng^{o_1=45, o_2=90}\). The red square represents the most active angle cell.
		\subref{fig:perception_res:arc} Activity of arc cells. The red square represents the most active arc cell. The overall excitation distribution shows continuity.
	}
	\label{fig:perception_res}
\end{figure}

	When running the model, $Orangutan$ reads the image data at the start of each tick, amplifies it by 100 times (0-25500) to serve as the excitation value for the pixel cells. The pixel cells consequently generate a certain spike frequency and transmit the signal forward to drive the operation of the entire neural network. Figure \ref{fig:perception_res:pixel} displays the activity level of the pixel cell layer after inputting an Mnist image (2\_72.bmp) into the model. The activity level of neurons in the figure accurately reflects the grayscale of pixels in the image.
	
	\begin{table*}[ht]
	\centering
	\caption{Brain regions in the model. The distribution of cortical columns is described in the format of hypercolumn rows $\times$ hypercolumn columns $\times$ minicolumns within each hypercolumn. The superscripts of neuron symbols correspond to different abstract attribute values in the Listing \ref{lst:vars_and_enums} ($o/o1/o2:ORIENTS,s:RECEPTIVE\_FIELD\_LEVELS,a:ANGLES$).The number of neurons in each minicolumn is obtained by multiplying the quantities of these attributes.}
	\label{table:regions}
	\small
	\setlength{\tabcolsep}{4pt}
	\renewcommand{\arraystretch}{1.3}
	\begin{tabular}{
			>{\centering\arraybackslash}m{2.3cm}
			|>{\centering\arraybackslash}m{2cm}
			|>{\centering\arraybackslash}m{2cm}
			|>{\centering\arraybackslash}m{1.7cm}
			|>{\centering\arraybackslash}m{2cm}
			|m{6cm}
		}
		\hline
		Brain Region & Column Distribution & Primary Neurons & Neuron Symbols & Neurons per Minicolumn & Neuron Description \\
		\hline
		\multirow{6}{*}{Point} & \multirow{6}{*}{$28\times28\times1$} & Pixel Cell & $CPxl$ & 1 & Represents a pixel point. \\
		\cline{3-6} && Edge Point Cell & \(CEdg^o\) & 16 & Extracts directed edge points. \\
		\cline{3-6} && Edge Point Feedback Cell &  N/A & 16 & Receives feedback from orientation cells to regulate edge point stimulation of orientation cells. \\
		\hline
		\multirow{3}{*}{Orients} & \multirow{3}{*}{$28\times28\times1$} & Orientation Cell & \(COri^{s,o,a}\) & 704 & Extracts edge points in a specific orientation at this location. \\
		\cline{3-6} && Orientation Feedback Cell & N/A & 704 & Receives feedback from antipodal or vector cells for downstream transmission. \\
		\hline
		\multirow{3}{2.3cm}{\centering Antipodal Points} & \multirow{3}{*}{$28\times28\times1$} & Antipodal Point Cell & \(CAntP^{s,o,a}\) & 704 & Extracts and represents antipodal point features. \\
		\cline{3-6} && Antipodal Feedback Cell & N/A & 704 & Receives feedback excitation from the winner arc cells and transmits downstream. \\
		\hline
		\multirow{3}{*}{Vectors} & \multirow{3}{*}{$28\times28\times1$} & Vector Cell & \(CVec^{s,o,a}\) & 704 & Extracts and represents vector features. \\
		\cline{3-6} && Vector Feedback Cell & N/A & 704 & Receives feedback excitation from winning angle cells and transmits downstream. \\
		\hline
		\multirow{4}{*}{Angles} & \multirow{4}{*}{$28\times28\times1$} 
		& Angle Cell & \(CAng^{o1,o2}\) & 256 & Extracts and represents angular features. \\
		\cline{3-6} && Attention Winner Cell & $CWin$ & 256 & Represents winning angular features in attention competition, transmitting feedback to lower-order features. \\
		\hline
		\multirow{4}{*}{Arcs} & \multirow{4}{*}{$28\times28\times1$} & Arc Cell & \(CArc\) & 1 & Extracts and represents arc features . \\
		\cline{3-6} && Attention Winner Cell & $CWin$ & 1 & Represents winning arc features in attention competition, transmitting feedback to lower-order features. \\
		\hline
		Current Position & \multirow{1}{*}{$1\times2\times1$} & Current Position Cell & $CCurX^{c}$ $CCurY^{c}$ & 28 & Coordinates for the focused features, used to direct eye movement. The superscript (c) in the symbol represents coordinates with values ranging from 0 to 28. \\
		\hline
		Previous Position & \multirow{1}{*}{$1\times2\times1$} & Previous Position Cell & $CPrevX^{c}$ $CPrevY^{c}$ & 28 & Records and represents the coordinates of the location of the previously noted feature. \\
		\hline
		Relative Position & \multirow{1}{*}{$1\times2\times1$} & Relative Position Cell & $CRelX^{c}$ $CRelY^{c}$ & 57 & Represents the position of the currently noticed feature relative to the previously noticed feature. \\
		\hline
		\multirow{11}{2.3cm}{\centering Abstract Attributes} & \multirow{11}{*}{$1\times1\times1$} & Relative Direction Cell & N/A & 16 & Represents the displacement direction of the current feature relative to the previous feature. \\
		\cline{3-6} && Relative Distance Cell & N/A & 11 & Represents the displacement distance of the current feature relative to the previous feature. \\
		\cline{3-6} && Type Cell & N/A & 2 & Represents the type of feature. With two enumeration values: angle and arc. \\
		\cline{3-6} && Orientation Cell & N/A & 17 & Represents the feature orientation. Includes 16 orientations from ORIENTS and one non-oriented. \\
		\cline{3-6} && Angle Cell & N/A & 9 & Represents the opening angle of the feature. Includes 8 angles from ANGLES and one non-angled. \\
		\cline{3-6} && Scale Cell & N/A & 11 & Represents the scale of the feature. \\
		\hline
		Global Regulation  & $1\times1\times1$ & Attention Competition Cell & $CComp$ & 1 & In the attention competition network, aggregates the excitation of all participating neurons and, in turn, applies an all-or-nothing inhibition to them. \\
		\hline
	\end{tabular}
\end{table*}
	
	\subsubsection{Extracting Edge Points}
	
	The two types of top-level features defined in this paper are located at the contours' edges, as shown in Figure \ref{fig:feature}. Edge points, serving as the foundation forming the contour edges, are the primary objects extracted by this model. An edge point is a directed point where the brightness of the pixel at its position is lower than that of its adjacent pixels. This model defines 16 orientation variables to describe the orientations of edge points within a polar coordinate system where the upward direction represents 0 or 360 degrees, with angles increasing in a clockwise direction, as seen in the $ORIENTS$ variable in Listing \ref{lst:vars_and_enums}. Correspondingly, within each edge column, there are 16 edge cells $CEdg^{o}$ (see Table \ref{table:regions}), where the superscript $o$ indicates the cell's selectivity for specific edge orientations (symbols for other types of feature cells follow the same format). The more prominent the related edge points in an image, the more active they are. Other orientations that lie between these 16 orientations are jointly represented by varying levels of activity in two adjacent edge cells.
	
\begin{figure*}[!t]
	\centering
	\subfigure[] {\includegraphics[width=0.48\linewidth]{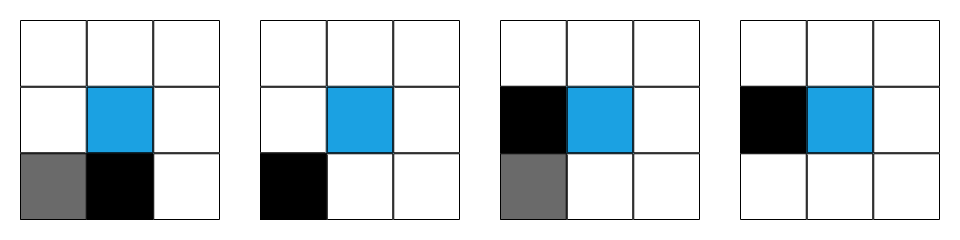} \label{fig:receptive_field:edge_point}}
	\hfill
	\subfigure[] {\includegraphics[width=0.48\linewidth]{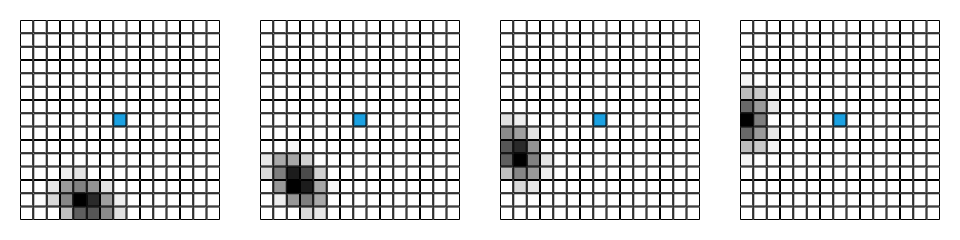} \label{fig:receptive_field:orient}}
	\hfill
	\subfigure[] {\includegraphics[width=0.48\linewidth]{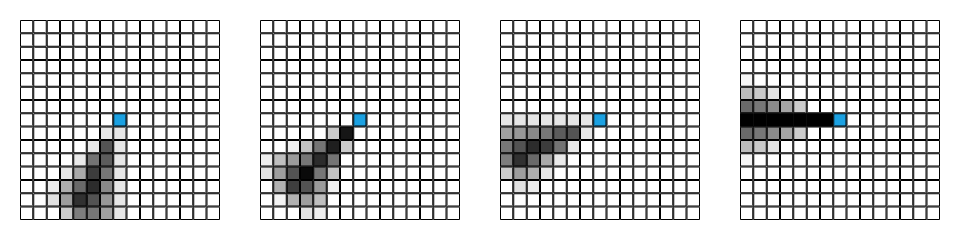} \label{fig:receptive_field:vector}}
	\hfill
	\subfigure[] {\includegraphics[width=0.48\linewidth]{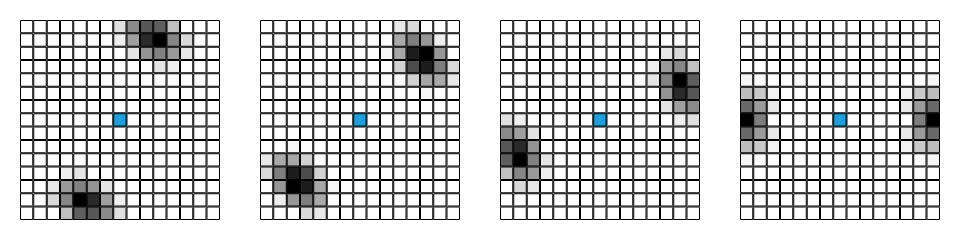} \label{fig:receptive_field:line}}
	\hfill
	\subfigure[] {\includegraphics[width=0.48\linewidth]{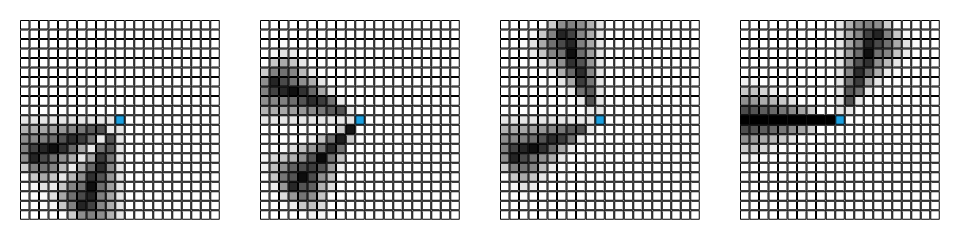} \label{fig:receptive_field:angle}}
	\subfigure[] {\includegraphics[width=0.48\linewidth]{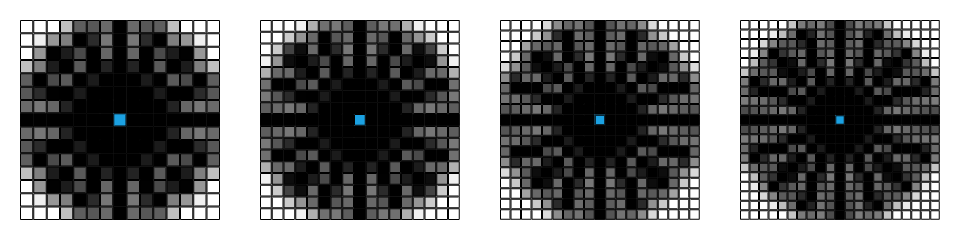} \label{fig:receptive_field:arc}}
	
	\caption{ Peripheral receptive fields of various feature cells. Blue squares represent the cell locations, and gray squares denote peripheral receptive fields, with the grayscale intensity roughly reflecting the synaptic strength established between the feature cells and the cells within the receptive field. For simplicity, this diagram only displays a subset of the receptive fields.
		\subref{fig:receptive_field:edge_point} Receptive fields of some edge point cells, adjacent to the cell location.
		\subref{fig:receptive_field:orient} Receptive fields of some orientation cells, which extend further from the center compared to edge point cells.
		\subref{fig:receptive_field:vector} Receptive fields of some vector cells, consisting of receptive fields from multiple orientation cells with the same orientation but sequential scales.
		\subref{fig:receptive_field:line} Receptive fields of some antipodal point cells, formed by receptive fields from two scale-matched, centrally symmetric orientation cells.
		\subref{fig:receptive_field:angle} Receptive fields of some angle cells, comprising all scaled vector cells in two specific orientations.
		\subref{fig:receptive_field:arc} Receptive fields of some arc cells, composed of receptive fields from all orientations and scales of antipodal point cells.
	}
	\label{fig:receptive_field}
\end{figure*}

	In order to extract edge points from images, this model simulates the response characteristics of Lateral Geniculate Nucleus (LGN) neurons to local contrast with their center-surround antagonistic receptive field structure \cite{Valois1982SpatialFS} and the V1 cortex's orientation-specific selectivity for edges \cite{Hubel1962ReceptiveFB}. The receptive field refers to the area in the environment whose stimuli can influence the electrical activity of a neuron. For neurons in the visual system, the receptive field specifically denotes the area on the retina where light stimuli can affect the neuron's response. This model has designed central and surrounding receptive fields for edge cells oriented in various directions, as illustrated in Figure \ref{fig:receptive_field:edge_point}. Based on the receptive fields, the model established a feed-forward circuit for edge cells. As shown in Figure \ref{fig:perception:edge_circuit}, a $CEdg^{o=135}$ receives excitation from pixel cells in the surrounding receptive field while also being inhibited by pixel cells in the central receptive field. The $CEdg^{o=135}$ becomes active when the excitation of its own position's $CPxl$ is less than the weighted sum of the excitation of $CPxl$ at adjacent positions. The excitation received by the edge cell follows Equation \ref{eq:edge_point}, where $n$ epresents the number of $CPxl$ in the perimeter receptive field, and subject to the condition that $\sum_{i=1}^{n} TRSPxl_j = 65$. In the equation, I replaced the letter ``C'' at the beginning of the symbols $CEdg^{o}$ and $CPxl$ with the symbols of attributes in the Table \ref{table:regions} to represent the attribute variables of the corresponding neurons. The formulas presented hereafter all adhere to this same format. Figure \ref{fig:perception_res:edge_point} showcases the activity of $CEdg^{o=135}$ on image $2\_72$.
	
	\begin{equation}
		EEdg^{o=\theta}_i = \sum_{j=1}^{n} TRSPxl_jEPxl_j - EPxl_i
		\label{eq:edge_point}
	\end{equation}
	
	\subsubsection{Extracting the Orientation}
	
	The receptive fields of edge point cells are relatively small, whereas corners and lines, as top-level features, have larger scale receptive fields. To allow small-scale features to represent large-scale ones, this model established orientation brain regions and orientation cells, as seen in Table \ref{table:regions}. As shown in Figure \ref{fig:receptive_field:orient}, orientation cells have receptive fields that extend further than those of edge point cells, denoted as $COri^{s,o,a}$, where $s$ denotes the scale (as shown in $RECEPTIVE\_FIELD\_LEVELS$ within Listing \ref{lst:vars_and_enums}) of the distance between the peripheral receptive field and the central receptive field in pixel units, $o$ denotes the orientation of the peripheral receptive field relative to the central receptive field, and $a$ represents the angle (as shown in $ANGLES$ within Listing \ref{lst:vars_and_enums}) between the direction $o1$ indicated by $CEdg^{o1}$ within the receptive field and the direction $o$ inherent to $COri^{s,o,a}$ itself. The receptive field of an orientation cell is a sector area defined by its scale and orientation, with the area being directly proportional to the scale. This ensures that no potential edge points are missed during the extraction process. Additionally, it should be noted that the receptive field scale defined by this model is the diameter of a circle with its center at the position of the neuron, obtained through the formula $s = 2n + 1$, where $n = 1, 2, \dots, 11$.
	
	Each $COri^{s,o,a}$ receives excitations from all specific-oriented $CEdg^o$ within its receptive field and takes the maximum value among them as its own excitation, as seen in Equation \ref{eq:feature_orient}, where $n$ is the total number of edge point cells within the receptive field. It establishes synapses with edge point cells at different locations, each having a different $TRS$, with those closer to the center of the receptive field having a higher $TRS$, and vice versa. The model primarily stimulates orientation cells of four types: $a=90/180/270/360$, which are used for subsequent feature extraction tasks. Figure \ref{fig:perception_res:orient} demonstrates the activity of $COri^{s=9,o=45,a=90}$ on image 2\_72.
	
	\begin{equation}
		EOri^{o,s,a}=\max\limits_{1 \leq i \leq n}TRSEdg^{o'=o+a}_iEEdg^{o'=o+a}_i
		\label{eq:feature_orient}
	\end{equation}
	
	The mechanism by which orientation cells integrate neuronal excitation within a region bears certain similarities to complex cells in neuroscience. Simple cells, such as edge point cells, are highly sensitive to specific visual stimuli and their locations. In contrast, complex cells, while sensitive to stimulus characteristics, are less precise about the exact location. This means that complex cells will respond as long as the stimulus attributes are correct, regardless of where the stimulus appears within the receptive field \cite{Hubel1962ReceptiveFB,Hubel1968ReceptiveFA}. Subsequent extraction and representation of more complex features based on orientation cells demonstrate robust responsiveness to the precise location of signals.
	
	\subsubsection{Extracting Vectors}
	
\begin{figure*}[!t]
	\centering
	\subfigure[] {\includegraphics[height=2.3in]{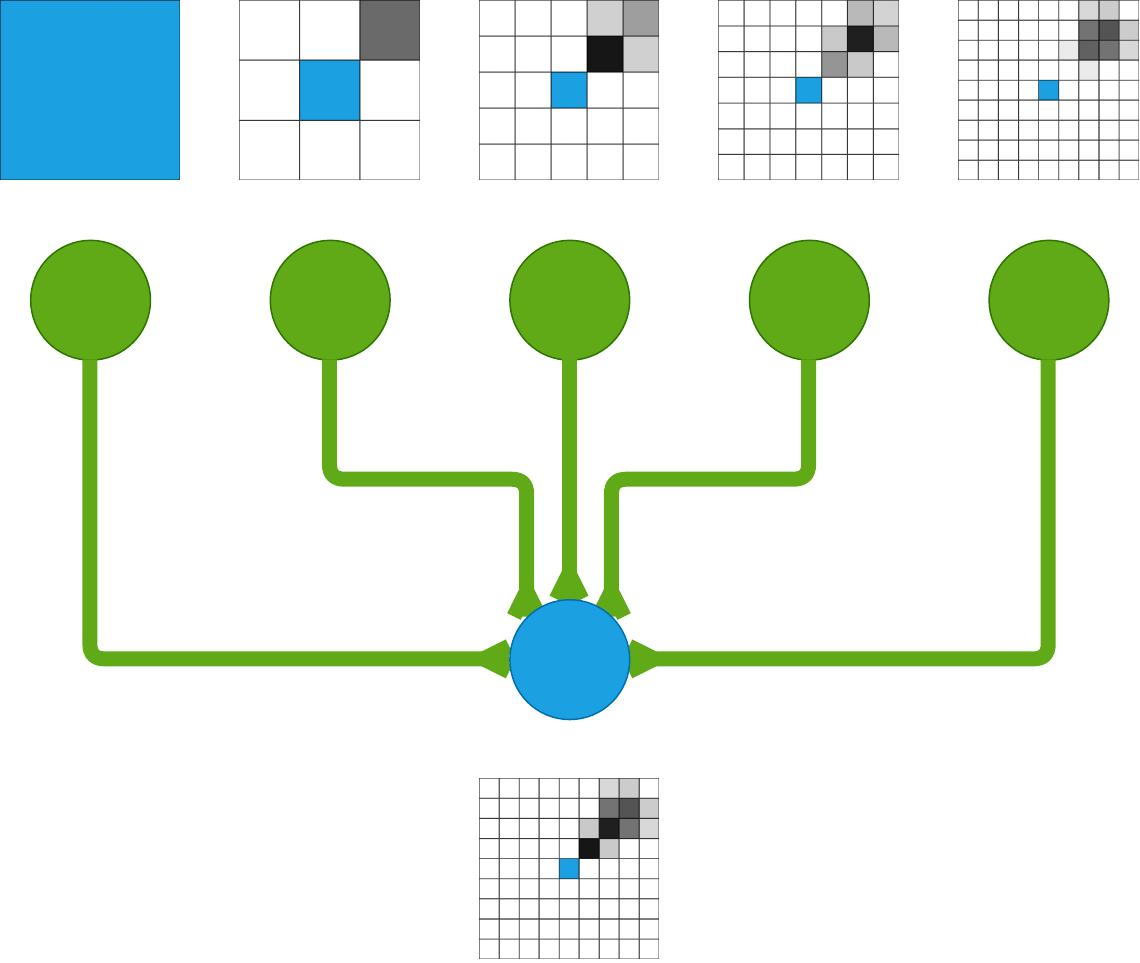} \label{fig:perception:seg_receptive_field}}
	\hfill
	\subfigure[] {\includegraphics[height=2.3in]{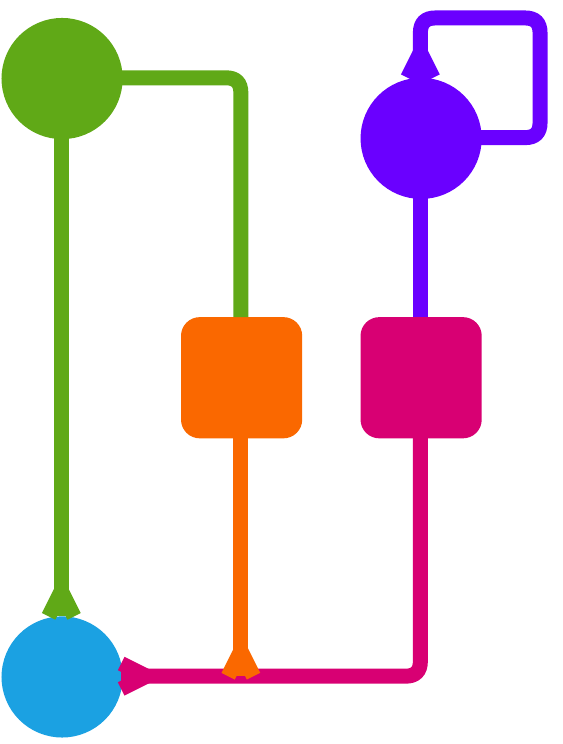} \label{fig:seg_perception:inhibit}}
	\hfill
	\subfigure[] {\includegraphics[height=2.3in]{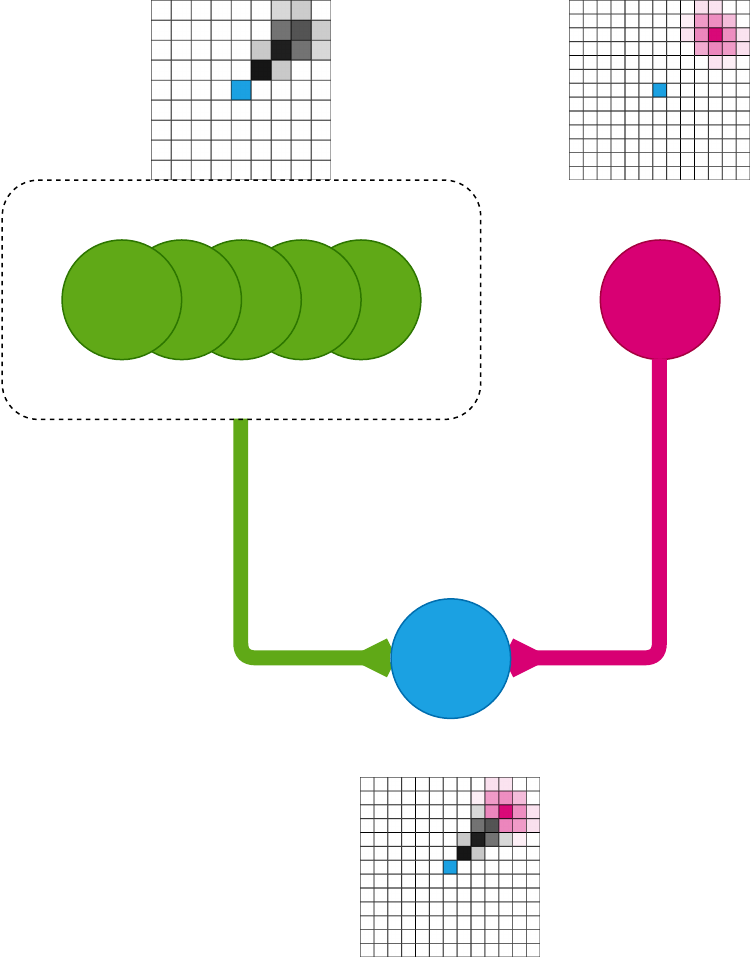} \label{fig:seg_perception:end_inhibit}}
	\hfill
	\caption{Feature extraction circuit for \(CVec^{s=9,o=45,a=90}\):
		\subref{fig:perception:seg_receptive_field} Five $COri^{s=1/3/5/7/9,o=45,a=90}$ (green) at the same location and across contiguous scales form excitatory synapses with a \(CVec^{s=9,o=45,a=90}\) (blue) at the same location, together constituting its receptive field.
		\subref{fig:seg_perception:inhibit} Penalty circuit during signal absence. Every orientation cell (green) is paired with a corresponding autaptic neuron (purple) that consistently transmits inhibitory excitation to the vector cell through inhibitory axons (red). Orientation cell counteracts this inhibitory effect by establishing a lateral inhibition circuit (orange) with it.
		\subref{fig:seg_perception:end_inhibit} End-inhibition circuit for \(CVec^{s=9,o=45,a=90}\). Larger scale \(COri^{s=13,o=45,a=90}\) (red) constitute the inhibitory receptive field for \(CVec^{s=9,o=45,a=90}\), which gets inhibited when feature signals cover this receptive field.
	}
\end{figure*}

	Vectors are defined as a specific type of linear structure, formed by a continuous arrangement of edge points, with a clearly designated start and end point. Like complex cells, vector cells  (denoted as $CVec^{s,o,a}$) integrate excitatory inputs from $COri^{s,o,a}$, which have the same orientation ($o$) and angle ($a$) across a continuous scale ($s$), to construct a perception of contour lines. The receptive fields of these cells are illustrated in Figure \ref{fig:receptive_field:vector}. This process follows the excitation formula outlined in Equation \ref{eq:extract_seg_1}, where $i=1, 3, 5, \ldots$. In Figure \ref{fig:perception:seg_receptive_field}, I present a neural circuit diagram illustrating the formation of a receptive field for $CVec^{s=9,o=45,a=90}$. When a vector feature of scale 9, fully matching this receptive field, appears, the corresponding neuron is maximally activated.
	
	\begin{figure*}[!htb]
		\begin{equation}
			ESeg^{s=n, o=m, a=\alpha } = \sum_{i=1}^{n}{EOri^{s=i, o=m, a=\alpha}}
			\label{eq:extract_seg_1}
		\end{equation}
	\end{figure*}
	
	\begin{figure*}[!htb]
		\begin{equation}
			ESeg^{s=n, o=m, a=\alpha } = \sum_{i=1}^{n}{(EOri^{s=i, o=m, a=\alpha} - \max(I-EOri^{s=i, o=m, a=\alpha},0))}
			\label{eq:extract_seg_2}
		\end{equation}
	\end{figure*}
	
	\begin{figure*}[!htb]
		\begin{equation}
			ESeg^{s=n, o=m, a=\alpha } = \sum_{i=1}^{n}{(EOri^{s=i, o=m, a=\alpha} - \max(I-EOri^{s=i, o=m, a=\alpha},0))} - (\lfloor n/2 \rfloor+1)EOri^{s=n+4, o=m, a=\alpha}
			\label{eq:extract_seg_3}
		\end{equation}
	\end{figure*}

	The method of activating vector cells through the simple linear addition of orientation cell excitation may lead to a problem: larger scale vector cells might receive more excitation signals more easily compared to smaller scale vector cells, even if their receptive fields are not fully covered by visual signals. This situation could reduce the precision of neuronal signal encoding. To address this issue, the model establishes an additional antagonistic inhibition circuit, as shown in Figure \ref{fig:seg_perception:inhibit}. This means that if there is any presynaptic input missing, the vector cell will suffer from additional inhibition, which is proportional to the degree of input missing: the more missing the input, the stronger the inhibition. Only when all types of $COri^{s,o,a}$ are active can the $CVec^{s,o,a}$ completely avoid such inhibition. Accordingly, I have updated the excitation equation to form a new formula \ref{eq:extract_seg_2}, where the constant $I$ represents the theoretical extremum of $EOri^{s=i, o=m, a=\alpha}$.
	
	In another scenario, the improper activation of vector cells can be attributed to large-scale vector features. When these large-scale features activate the vector cells of the corresponding scale, they might also completely cover the receptive fields of smaller scale vector cells, erroneously causing these cells to be fully activated. To address this issue, this model introduced a method simulating the end-inhibition mechanism found in the visual cortex \cite{Bolz1986GenerationOE}. As shown in Figure \ref{fig:seg_perception:end_inhibit}, $COri^{s=13, o=45, a=90}$ forms an inhibitory synaptic connection with $CVec^{s=9, o=45, a=90}$. Given that $COri^{s=13, o=45, a=90}$ is located on the periphery of $CVec^{s=9, o=45, a=90}$'s receptive field, this means when a vector feature extends beyond the boundary of the receptive field, $COri^{s=13, o=45, a=90}$ gets activated and sends an inhibitory signal to $CVec^{s=9, o=45, a=90}$, thereby inhibiting its activity. It is worth noting that in this model, the receptive fields involved in end-inhibition are 4 scales larger than the excitation receptive fields. This design prevents the end-inhibition response from being overly sensitive and mistakenly activated due to the inhibitory and excitatory receptive fields being too close together. The formula for calculating excitation has been updated to Equation \ref{eq:extract_seg_3}. This end-inhibition mechanism ensures that vector cells are fully activated only when the vector features properly cover their receptive fields, thereby enhancing the precision of vector cell encoding. Figure \ref{fig:perception_res:seg} shows the activity of $CVec^{s=9, o=45, a=90}$ in image 2\_72.
	
	\subsubsection{Extracting Angles}
	
	In this model, an angle is defined as being comprised of two vectors, constrained to be between 45 degrees and 180 degrees, segmented in steps of 22.5 degrees, as shown in $ANGLES$ within Listing \ref{lst:vars_and_enums}. To reduce the complexity of the algorithm, the model further simplifies the definition of an angle to be scale-invariant, meaning that vectors of any scale in the two directions can be extracted as feature angles. Previous research indicates that in the primary visual cortex of the brain, there exist specific neurons that selectively respond to angles and exhibit the highest activity when the apex of an angle is located at the center of their receptive fields \cite{Tang2018ComplexPS}. Based on this finding, this model designed angle cells (denoted as $CAng^{o1,o2}$) to simulate the functions of such neurons. The excitation formula for angle cells, as described in Equation \ref{eq:feature_angle}, reflects a higher activation level of the angle cell when the excitations of two vectors at specified orientations are similar, where $n=1, 3, 5, \ldots$. Additionally, the formula introduces a constant $\lambda$ (set to the value of 0.08) to bolster the excitation of $CAng^{o1,o2}$ at smaller angles, aligning the feature extraction results more closely with the human intuitive perception of object angles. Figure \ref{fig:perception_res:ang} demonstrates the active state of the angle cell $CAng^{o_1=45, o_2=90}$ in image 2\_72.
	
	\begin{figure*}
		\begin{multline}
			EAng^{o1=i, o2=j} = 65(1 - \cfrac{\lambda |i-j|}{22.5})\min(\max\limits_{1 \leq n \leq 21}EVec(s=n, o=i, a=90),\max\limits_{1 \leq n \leq 21}EVec(s=n, o=j, a=270))
			\label{eq:feature_angle}
		\end{multline}
	\end{figure*}
	
	\subsubsection{Extracting Arcs}
	
	In this model, arcs are defined as contour edges in the image with a specific curvature. Similar to the processing of angle features, according to related research, there are neurons in the primary visual cortex that are specifically activated when the apex of a contour is located at the center of their receptive fields \cite{Tang2018ComplexPS}. In addition to arc features, the frequently occurring ring features in images also need to be effectively extracted and represented. To construct a unified algorithmic framework to handle both types of features, this model considers arcs and rings as the same category of feature, adopting a different approach from the literature to define the receptive field of arc cells by setting the center of the receptive field at the centroid of the arc or ring. Furthermore, given the presence of numerous irregular, approximately arc-shaped curves in images, including those with distinct edges, this model also includes these irregular curves within the scope of arc features and extracts them using a unified algorithm. Arc cells are denoted as $CArc$ and are insensitive to scale and opening orientation.
	
	Similar to feature angles, the extraction of arc features also follows a hierarchical process from simple to complex. Initially, the receptive fields of two orientation cells $COri^{s,o,a}$ with opposite directions and the same scale jointly constitute the receptive field for antipodal point cells (denoted as $CAntP^{s,o}$), as shown in Figure \ref{fig:receptive_field:line}. This design follows an excitation formula \ref{eq:antipodal_percept} to establish a feed-forward neural circuit with $CAntP^{s,o}$, where $TRS(s)$ represents larger excitation from $COri^{s,o,a}$ with larger scales. Subsequently, $CAntP^{s,o}$ of various orientations and scales together form the receptive field for $CArc$, as presented in Figure \ref{fig:receptive_field:arc}, and a feed-forward neural circuit is established with $CArc$ according to the excitation formula \ref{eq:arc_percept}. Antipodal points ensure symmetry during the extraction of arc features, while the aggregation of antipodal points of different scales in the same orientation increases the robustness of the extraction algorithm for non-regular arc features. Considering the similarity between the extraction processes and principles of arc- and angle-related features, this paper does not elaborate on the specific neuro-network structures and detailed descriptions of arc feature extraction. Figure \ref{fig:perception_res:arc} illustrates the activation of arc cells in image 2\_72.
	
	\begin{figure*}
		\begin{align}
			EAntP^{s=\theta,o=\omega} = TRS\min(EOri^{s=\theta,o=\omega, a=180}, EOri^{s=\theta,o=(\omega+180)\mod360, a=180}) && \text{where $TRS = 1.2 + 0.15\theta$}
			\label{eq:antipodal_percept}
		\end{align}
	\end{figure*}
	
	\begin{figure*}
		\begin{align}
			EArc = \sum_{i=1}^{8}{TRS\max\limits_{0 \leq n \leq 21}EAntP^{s=n,o=22.5i}} && \text{where $TRS = 1 / 8$ and $n=1,2,\dots$}
			\label{eq:arc_percept}
		\end{align}
	\end{figure*}
	
	\subsection{Motion}
	
	The saccade function is a key motor process in this model, primarily comprising three stages: attentional competition, eye saccades, and attention shift, as depicted in the blue part of Figure \ref{fig:model_workflow}.
	
	\subsubsection{Attentional Competition}
	
	During the perception phase, this model generates a saliency map \cite{Koch1985ShiftsIS} concerning features such as angles and arcs, where the activity levels of angle cells and arc cells represent the physical saliency of stimuli in the visual field. Subsequently, the model needs to filter out a single, most active cell from all angle and arc cells, which will serve as the target for the next step of the eye saccade. To this end, a winner-takes-all (WTA) competitive relationship is established among the feature cells, such that the final victor maintains full activation while the losers become completely inactive. According to Koch et al., a WTA neural network is equivalent to a maximum value operator \cite{Koch1985ShiftsIS}, which corroborates the working principle of the winner-takes-all mechanism in this model: only the strongest signal source will stand out in a competitive environment.
	
\begin{figure}
	\centering
	\includegraphics[width=0.6\linewidth]{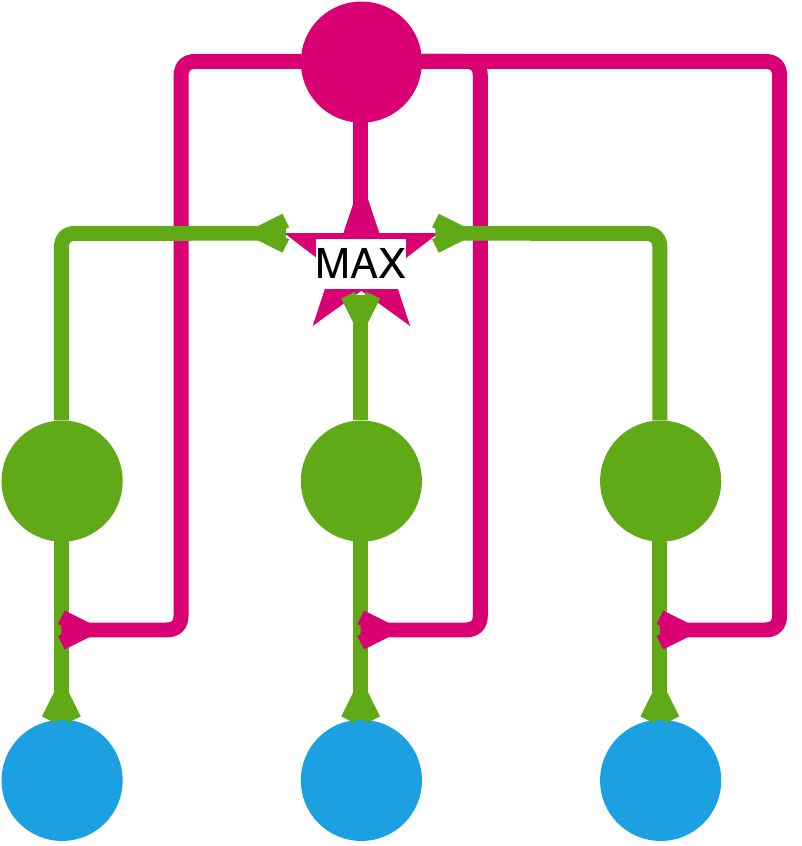}
	\caption{Attention Competition Circuit. Feature cells (green) involved in the competition form excitatory synapses with their respective winner cells (blue). The attention competition cell dendrites (red) receive excitation from all the feature cells, compute the maximum value, and prevents the excitation process of feature cells towards winner cells through all-or-nothing weak inhibitory synapses (red). }
	\label{fig:attentioncompetitioncircuit}
\end{figure}

	This model establishes a centralized WTA network to implement the mechanisms described, as shown in figure \ref{fig:attentioncompetitioncircuit}. Initially, an ``Attention Competition Cell'' (denoted as $CComp$) aggregates the excitation of all $CAng$ and $CArc$ through its MAX dendrite. Concurrently, each feature cell is paired with a corresponding winner cell located within the same column. Feature cells must excite their associated winner cells to indicate their victory in attention competition. However, $CComp$ counters this by establishing an all-or-nothing weak inhibitory synapse ($all\_or\_none=1$), aiming to obstruct the process. Due to the nature of this synaptic connection, only the most active feature cell, whose excitation precisely equals the inhibitory excitation, can activate its winner cell without being affected by the inhibition. This effectively suppresses less active cells from similar activation, ensuring that each attention competition activates only a single winner cell.
	
	\subsubsection{Eye Saccades}
	
	Research suggests that attention fundamentally forecasts the destination of the next eye movement, implying that our focus on specific segments of the visual world typically triggers our eyes to move to that location \cite{ZHAO201240}. This model characterizes the specific location of the winning feature based on simulations of grid cells, serving as the target for eye movement. Grid cells are a class of neurons found in the brain that are capable of generating a regular hexagonal grid pattern across the entire environment. These patterns are believed to be part of the internal spatial mapping system of the brain, assisting animals in understanding their position within an environment \cite{Hafting2005MicrostructureOA}. Inspired by this, this model constructs a $28\times28$ grid cell matrix as a positional map and establishes a topological mapping relationship with the top-level features. When a top-level feature cell wins the attentional competition, its corresponding winner cell activates the grid cell at the relevant position. Due to the one-hot nature of the attentional competition mechanism, correspondingly, only one grid cell in the grid matrix will be active at any given time, serving as the sole target for the eye saccade.
	
\begin{figure}
	\centering
	\includegraphics[width=0.6\linewidth]{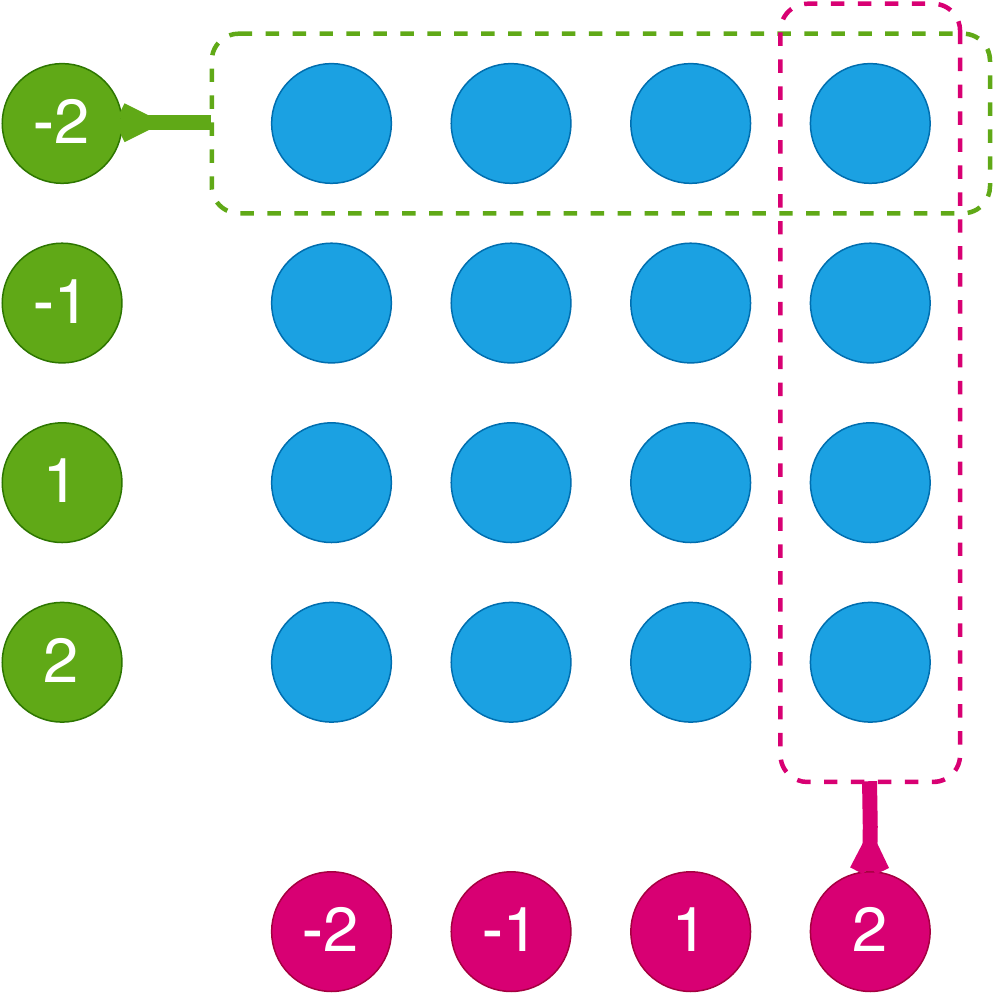}
	\caption{Grid cells (blue), representing feature locations, establish a Y-axis mapping (green dashed box) with neurons (green) controlling vertical eye movements, and an X-axis mapping (red dashed box) with neurons (red) controlling horizontal eye movements. The absolute value of numbers on the motor neurons represents the distance of eye movement control, with positive numbers indicating movement downward or to the right, and negative numbers indicating movement upward or to the left.}
	\label{fig:saccade}
\end{figure}

	Biologically, the motor cortex drives the process of eye saccades through neuromuscular junctions. The extraocular muscles around the eye are influenced by neural signals, jointly regulating the rotation angle of the eye \cite{doi:10.1152jn.1970.33.3.393}. This model establishes a set of positional neurons (denoted as $CCurX$ and $CCurY$) to represent the coordinates of the currently focused feature and to approximately simulate the input of eye movement signals. By connecting them with grid cells, a mapping relationship between target position and eye state can be established, as shown in Figure \ref{fig:saccade}. This mechanism can be applied in the field of robotics, to drive changes in the posture of cameras or motors. As this research focuses on the algorithmic level, the integration of software and hardware is not extensively discussed in this paper.
	
	Theoretically, eye saccades to a new position cause changes in input signals, involving complex visual processing mechanisms. However, to simplify the computational process, this study does not simulate this visual input change effect. Such simplification allows for a more focused exploration and validation of the underlying control mechanisms of eye movement, without the complexities of visual processing. Future work could further refine the model, including the effects of visual feedback on the eye control system, to achieve a more comprehensive brain-inspired artificial intelligence system.
	
	\subsubsection{Attention Shift}
	
	Saccades are a continuous process, meaning that attention shifts from one feature to another, such as from the angle shown by the red square in Figure \ref{fig:perception_res:ang} to the arc represented by the red square in Figure \ref{fig:perception_res:arc}.
	
\begin{figure}[!t]
	\centering
	\subfigure[] {
		\includegraphics[height=2in]{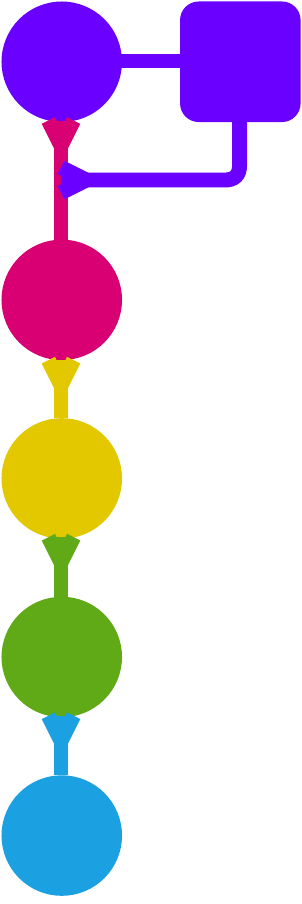} 
		\label{fig:attention_back_inhibitation:micro_scale}
	}
	\hfil
	\subfigure[] {
		\includegraphics[height=2in]{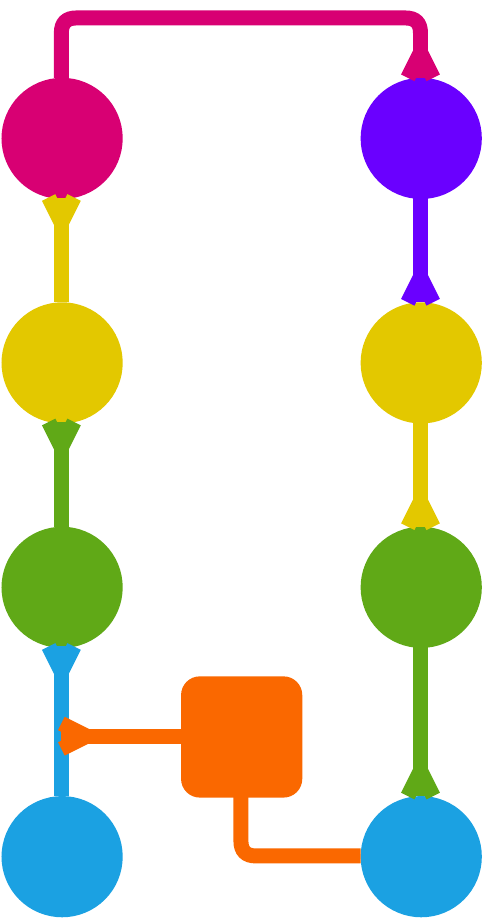} 
		\label{fig:attention_back_inhibitation:macro_scale}
	}
	\caption{Two neural circuits for attention shift through feedback inhibition. For simplicity, only the circuits related to feature angles are shown, with the feedback inhibition mechanism for feature arcs following a similar principle. \subref{fig:attention_back_inhibitation:micro_scale}. When input signals pass through $CEdg^{o}$ (blue), $COri^{s,o,a}$ (green), and $CVec^{s,o,a}$ (yellow) to reach $CAng^{o1,o2}$ and eventually activate the winner cell (purple) in the attention competition, winner cell applies recurrent inhibition to its own feedforward synapses through STD axon (purple square). \subref{fig:attention_back_inhibitation:macro_scale} winner cell (purple) conveys a negative feedback signal to the underlying $CEdg^{o}$ through a top-down, large-scale circuit established based on feedback cells (right path), which then applies STD effects (orange) to its feedforward circuit (left path). To simplify, although the actual network structure is much more complex than illustrated, every feature level is conceptually represented by a pair of neurons: one feedforward and one feedback. Additionally, since winner cells inherently possess receptive fields of the corresponding top-level feature cells, they can directly serve as the initiation points for the right-side feedback loop.}
\end{figure}

	A typical approach to implementing attention shifts is to establish a recurrent inhibitory circuit from the winner cell to its own feed-forward excitation circuit \cite{Koch1985ShiftsIS}. Influenced by its self-imposed inhibitory effect, the winner cell will reduce its activity level for a period of time, which is reported biologically to be 500ms \cite{Posner1982NeuralSC}, and in this models may be appropriately extended, for example, spanning the entire observational cycle for the current sample. This mechanism allows the competition network to sequentially produce different winners, as illustrated in Figure \ref{fig:attention_back_inhibitation:micro_scale}, where a winner cell temporarily inhibits its own activity, permitting other feature winner cell to become active in subsequent competitions and attract eye saccades.
	
	However, this mechanism has its limitations in this model: due to the population coding characteristic of neural networks, a feature will activate multiple neurons to varying degrees, and their activity distribution is continuous, as seen in Figure \ref{fig:perception_res:arc}. Therefore, by solely inhibiting the excitatory source of the winner cell, attention will shift to another nearby similar feature, making it difficult to notice other more novel features, hence reducing the observational efficiency of the model. A relatively direct solution to this is to inhibit all feature cells within a circular area centered on the winner cell on the saliency map, allowing attention to shift to features at a greater distance \cite{Itti2000ASS}. However, this introduces additional accuracy problems: if the inhibition range is too small, the desired effect will not be achieved, while too large a range may erroneously inhibit feature cells of high observational value. Furthermore, due to the diversity and complexity of real-world samples, it's challenging to find an optimal scale that applies to all possible scenarios.
	
	\begin{figure}[!t]
	\centering
	\subfigure[] {\includegraphics[height=1in]{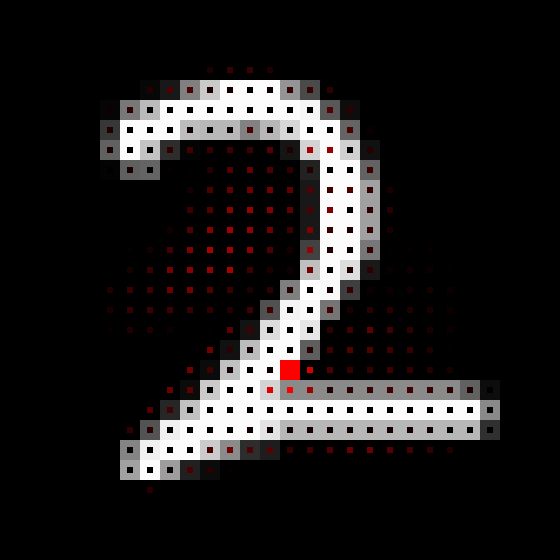} \label{fig:attention:winner}}
	\hfill
	\subfigure[] {\includegraphics[height=1in]{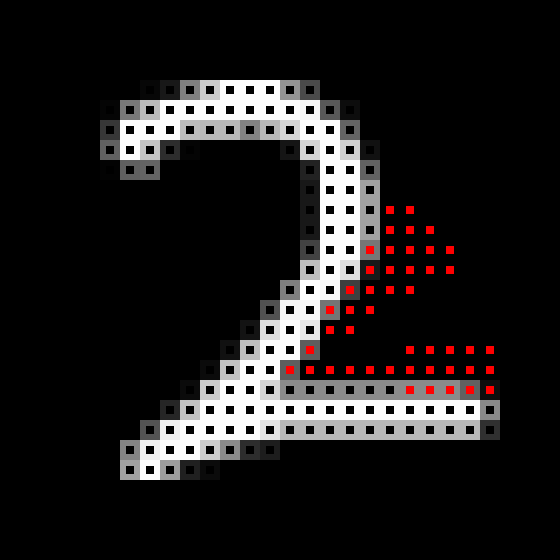} \label{fig:attention:feedback}}
	\hfill
	\subfigure[] {\includegraphics[height=1in]{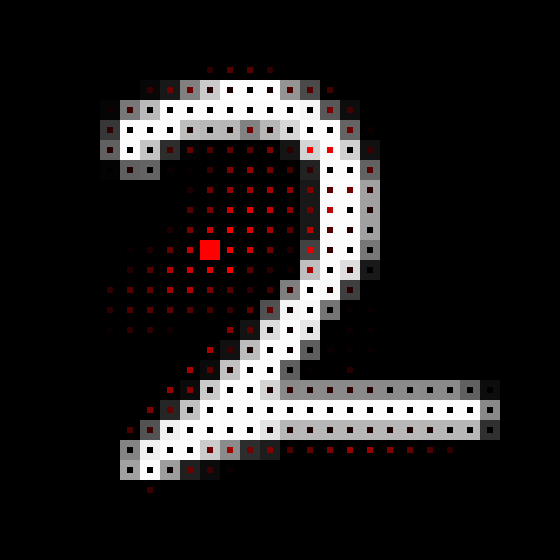} \label{fig:attention:feedback_inhibit}}
	\caption{Attention Competition
		\subref{fig:attention:winner} Among all active feature cells (red dots), an angle cell with the highest excitation (red square) wins the attention competition.
		\subref{fig:attention:feedback} The winning angle cell conveys feedback excitation to all feedback edge dot cells (red dots) within its receptive field.
		\subref{fig:attention:feedback_inhibit} All angle cells with receptive fields similar to that of the winning angle cell become inactive due to the loss of feedforward excitation, allowing an arc cell (red square) to become the new winner.}
	\label{fig:attention}
\end{figure}
	
	To address the issues above, this model establishes a top-down recurrent inhibitory circuit. This circuit shares the same trans-brain-area path as the bottom-up feature extraction circuit but operates in the opposite direction, as shown in Figure \ref{fig:attention_back_inhibitation:macro_scale}. This negative feedback pathway has excellent directionality, and ultimately, only the $CEdg^{o}$ of all orientations located within the receptive field of winner cell can obtain the negative feedback excitation. Therefore, the shape and range of the recurrent inhibitory area correspond to the actual receptive field of winner cells. Figure \ref{fig:attention:feedback} displays the activity of $CEdg^{o}$'s feedback cells excited by a $CAng^{o1=45,o2=90}$'s winner cell in the reverse direction on image 2\_72. Subject to feedback inhibition, $CEdg^{o}$ reduces its own excitatory transmission, affecting not only the winner feature cells themselves but also all $CAng^{o1,o2}$ and $CArc$ that receive excitation from these $CEdg^{o}$, to varying degrees. This approach balances precision and generalizability when executing negative feedback inhibition on feature cells, thereby enhancing the model's efficiency in recognizing novel and distinct features. Eventually, with $CAng^{o1=45,o2=90}$ inhibited, another $CArc$ successfully captures the model's attention, as depicted in Figure \ref{fig:attention:feedback_inhibit}.
	
	This top-down negative feedback regulatory process exhibits mechanisms analogous to those in the brain: visual signals from the retina are relayed through the lateral geniculate nucleus (LGN) and are modulated by the influence of the thalamic reticular nucleus (TRN) before being transmitted to the primary visual cortex (V1). The TRN receives excitatory back-projections from V1 and establishes inhibitory projections to the LGN \cite{McAlonan2006AttentionalMO}. This fact provides a biologically plausible source for the attention algorithm in this model.
	
	\subsection{Abstraction}
	
\begin{figure}
	\centering
	\includegraphics[width=0.5\linewidth]{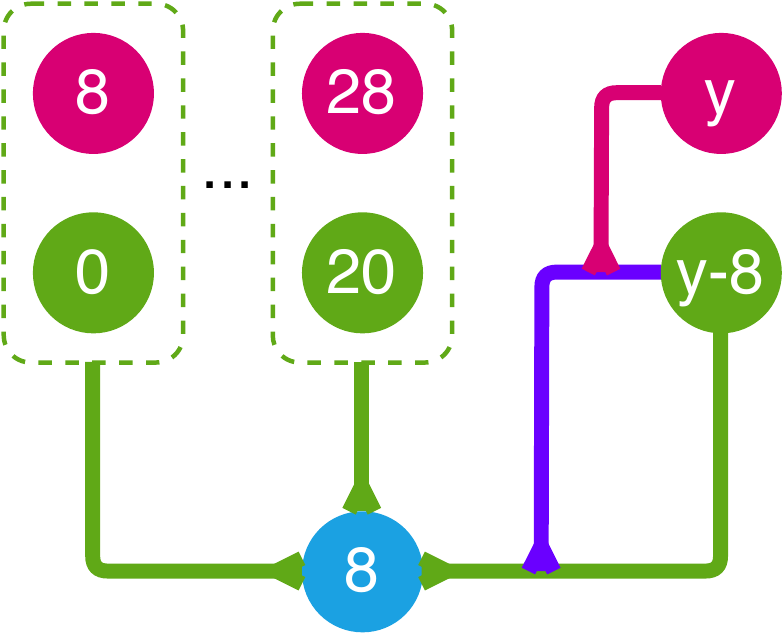}
	\caption{Feature Inter-Relative Position Perception Circuit. In the diagram, $CRelY^{c=8}$ (blue) represents a displacement of 8 pixels on the Y-axis and can be activated by various combinations (dashed boxes) of $CCurY$ (green) and $CPrevY$ (red), where each combination internally reflects a Y-axis coordinate difference of 8. On the far right, their joint stimulation of $CRelY^{c=8}$ through an AND gate neural microcircuit is depicted, with green lines indicating excitatory synapses and purple and red lines representing inhibitory synapses.}
	\label{fig:relative_pos}
\end{figure}

	Related studies have shown that during the process of continuous scanning, models establish the relative positional relationships between features \cite{Zimmermann2014BuildupOS}. In order to simulate this mechanism, I first established a set of coordinate cells, $CPrevY^{c}$ and $CPrevX^{c}$, that represent the position of the previous feature. Subsequently, before scanning to the next feature, the current feature position cells ($CCurY^{c}$ and $CCurX^{c}$) activate $CPrevY^{c}$ and $CPrevX^{c}$. Finally, the above two groups of neurons, through the neural circuitry as shown in Figure \ref{fig:relative_pos}, activate neurons $CRelY^{c}$ and $CRelX^{c}$ that represent the relative positions between features. Furthermore, through a simple AND gate logic circuit, this model can establish mapping relationships between $CRelY^{c}$ and $CRelX^{c}$ and neurons that represent abstract properties, the ``Relative Direction Cell'' and the ``Relative Distance Cell'', thereby extracting the relative direction and distance between two features. This abstraction capability for relative positions of features helps the model to establish a stable representation of the internal structure of objects, which is of significant importance for the learning capabilities and pattern recognition abilities of intelligent agents.
	
	In addition to locational properties, this model also extracts abstract attributes such as orientation, angle, and scale of top-level features. Taking scale as an example, I first established attribute cells representing 11 scales $CScl^s$. Then, using the characteristic that $CWin$, which has won the attentional competition, passes down feedback excitation from the top-down, I established a mapping circuit between the feedback cells of $CVec^{o=45,s=5,a=90}$ and $CAntP^{s,o}$ and $CScl^s$. Lastly, I set up additional neural circuits to enhance the algorithm's precision, such as establishing local competitive circuits to ensure that for each orientation, only one $CVec^{o=45,s=5,a=90}$ or $CAntP^{s,o}$ with the maximum excitation is used for attribute mapping calculations, as well as averaging the scales represented by different orientations of $CVec^{o=45,s=5,a=90}$ or $CAntP^{s,o}$ when they are not the same to determine the scale of the winning feature. The neural algorithmic implementation for extracting orientation and angle attributes also follows similar principles. Due to space constraints, the specific excitation formulas and neural circuit structures will not be elaborated on in the article.
	
	The above abstract attributes are significant for this study. On one hand, they complement the perceptual abilities of this model. On the other hand, they will serve as the foundation for a symbolized cognitive system, to be applied in the future development of cognitive functions, including learning, in the $Orangutan$ cognitive system.
	
	\section{Results}
	
	I composed a test set of 100 samples using the first 10 images of each digit from the MNIST dataset and validated the feature observation capability of the model with it. The main objective of the testing was to examine whether the model could observe salient features in the images and make saccadic movements between these features. Salient features are defined as those that provide a good generalization of the image content, typically featuring larger scale and receptive fields covering more edge points. The test focused on whether the features discovered by the model met the definition of salient features and the efficiency with which the model observed novel salient features. Based on the model's operational principle, the model would observe all potential features until the negative feedback mask covers all active edge point cells, thereby returning all feature neurons on the saliency map to a resting state. To improve test efficiency, I capped the number of features observed by the model at five. Figure \ref{fig:saccade_path} shows part of the test results, with the complete saccadic trajectory data available in Appendix \ref{sec:saccade_test_result}.
	
	\begin{figure*}
	\centering
	\includegraphics[width=1\linewidth]{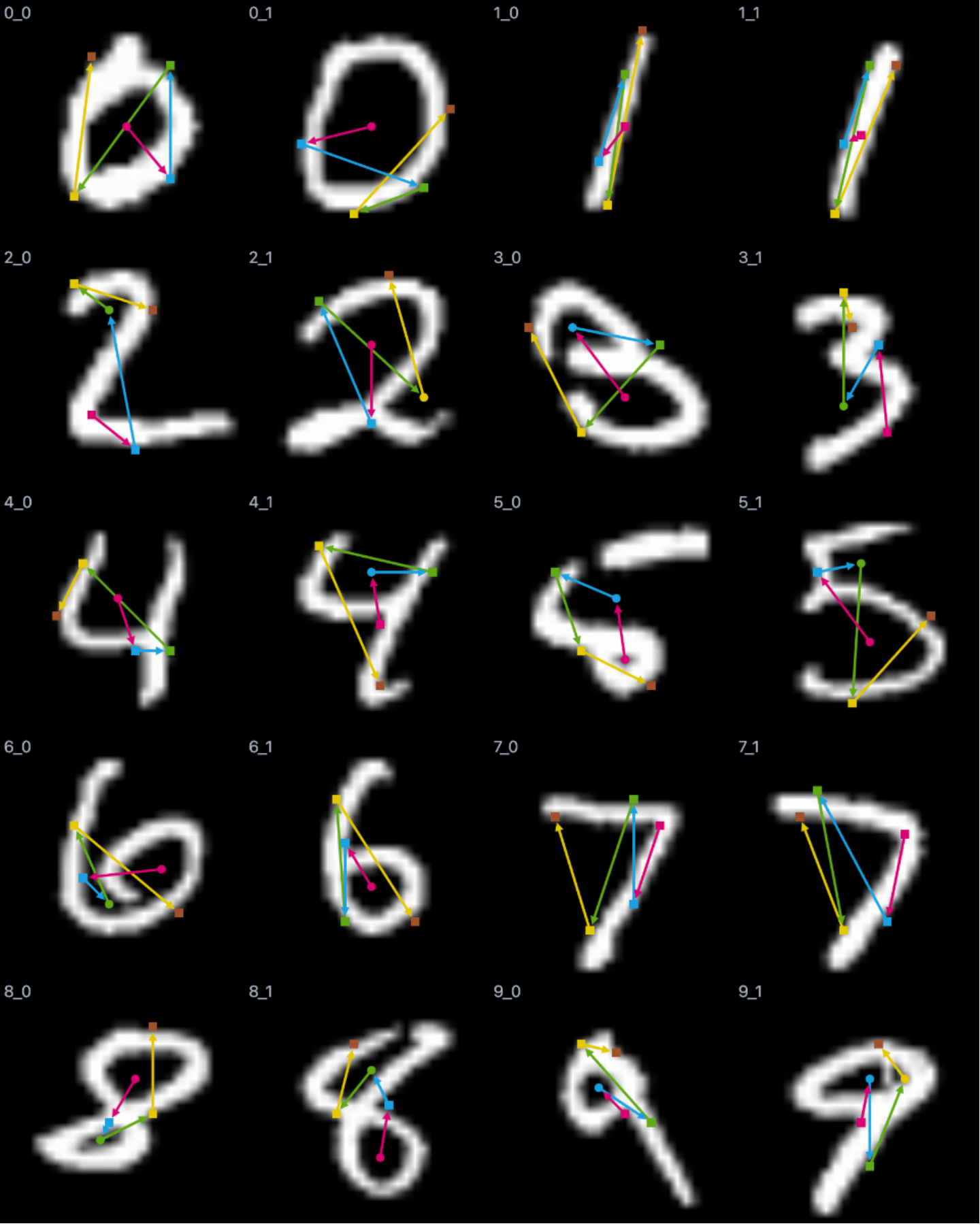}
	\caption{Partial Test Results Showcase for the Model's Scanning Function. In the picture, circles represent arcs, while squares denote angles. For further details on each feature, refer to the appendix \ref{sec:saccade_test_result}.}
	\label{fig:saccade_path}
\end{figure*}
	
	It should be noted that since the primary goal of this model is to simulate human eye saccade activity, no targeted testing was conducted for ancillary functionalities implemented in the algorithm, such as abstraction, and therefore related data is not presented in the test results. Additionally, the primary aim of this model is to demonstrate the effectiveness of the $Orangutan$ algorithm by showcasing its working principles, hence no further testing was conducted on a larger scale of samples.
	
	\section{Discuss}
	
	\subsection{Result Analysis}
	
	The results above indicate that the majority of features observed by the model on the test samples are consistent with the standards of salient features and possess sufficient novelty, generally meeting experimental expectations. However, some instances in the test results do not completely align with human expectations or intuition. First, the model extensively observed a category of sub-salient features in the samples of the digit ``1'', such as the second and third features of 1\_0 and the second feature of 1\_1. Although they possess larger receptive fields, they fail to cover the maximum number of edge points; therefore, their saliency is not the strongest, which deviates somewhat from human intuition. A similar issue also exists in other digit samples featuring large-scale linear edges, such as the fifth feature of 9\_0. The primary cause of this phenomenon is that handwritten digits are typically rough, and the serrated protrusions found in large-scale linear edges tend to produce small angular features locally. Coupled with the feature extraction algorithm's additional gain for small angles of $CAng^{o1,o2}$ (refer to Equation \ref{eq:feature_angle}), the model is more prone to focus on feature angles with smaller angles (which also tend to have smaller receptive field ranges). Additionally, larger-scale feature angles are more negatively affected by edge serration when exciting $CAng^{o1,o2}$. This issue could be resolved in future research with the establishment of new line cells and feature extraction neural algorithms specifically optimized for linear features. 
	
	Second, in samples of the digit ``7'', the model tends to notice small-scale feature angles at the ends of large-scale angles, such as the fourth and fifth features in 7\_0 and 7\_1. This issue also exists in other samples with large-scale angles, such as the fourth and fifth features of 1\_0 and 1\_1. Due to the model's receptive field scale being limited to 21, when the most salient $CAng^{o1,o2}$ is initially observed, its feedback receptive field is not broad enough to entirely cover the actual edges of feature angles in the image, allowing the areas not covered by the receptive field to be observed as independent features subsequently. This could be rectified by expanding the numerical range of scales.
	
	Finally, the model observes numerous non-core features that do not aid in digit recognition. This phenomenon is particularly evident in digit samples with fewer core features, such as ``0''. This happens because the current model algorithm only has a preference for feature saliency, tending to passively search for features that have not been observed previously. However, core features typically possess distinct characteristics of salient features, and the test results also demonstrate that the model tends to prioritize observing core features before peripheral ones, thus aligning with the design goals of the model.
	
	\subsection{Limitations and Future Directions}
	
	This model has achieved the goal of demonstrating the biological plausibility and algorithmic efficacy of $Orangutan$ through simulating the neural mechanisms of human eye saccades. However, there remain significant differences in terms of the final effectiveness as well as the size and complexity of the neural network compared to the biological brain. Here I outline the main limitations, which will be the focus of future research.
	
	\subsubsection{Incorporating More Brain-like Mechanisms}
	
	Learning is one of the core abilities of intelligent systems, and the current model essentially represents a static neural network based on a set of rule systems. Future plans include developing a learning mechanism centered around Hebbian rules. Leaning on learning mechanisms and feature position information and abstract attributes acquired during saccades, the model would dynamically establish new synapses and change network structures while observing. This would continuously build and update internal representations of digits, aiming to achieve human-level sustained learning capability.
	
	Furthermore, the existing mechanisms are still insufficient, as many excellent computational mechanisms in the brain have not yet been implemented. However, since $Orangutan$'s foundational implementation principles are close to those of the biological brain, it has good compatibility with new mechanisms. This allows the future introduction of more brain-like mechanisms in my research, moving towards the long-term goal of creating a general artificial intelligence system that can rival human capabilities.
	
	\subsubsection{Optimizing Model Algorithm}
	
	The current saliency-based attention mechanism within neuroscience is known as an involuntary attention process, which lacks conscious activity participation and does not possess the capability for active feature selection, making the model prone to observe some non-core features. Therefore, the next step is the proposed implementation of a voluntary attention mechanism: based on learning capabilities, the model could establish associations between different numbers and their corresponding core features. This would allow the model to generate preferences for specific features and actively search capabilities when observing specific numbers, thereby improving observation efficiency and robustness to noise.
	
	Additionally, in terms of feature representation, the current model only covers a limited range of feature types, such as angles and arcs, whereas the biological brain can recognize many more complex features, including lines and crosses \cite{Tang2018ComplexPS}. Future research will aim to increase the variety of feature extraction types in the model, to more comprehensively simulate the richness of human vision.
	
	Lastly, the current model has not been fully validated in terms of processing large-scale, complex-content, and low signal-to-noise ratio grayscale images and does not support color image processing. Future studies will expand the model's scope of application to accommodate a greater variety of visual environments and verify its performance on more complex and large-scale image samples.
	
	\subsection{Related Work}
	
	Classical neuron models such as Hodgkin-Huxley (HH), Integrate-and-Fire (IF), and Leaky Integrate-and-Fire (LIF) model various aspects of neurons transmitting signals in the form of spikes to different degrees, leading to the development of Spiking Neural Networks (SNNs) and Spike-Timing-Dependent Plasticity (STDP) learning algorithms. In comparison, $Orangutan$ has not yet achieved the same level of precision in simulating these electrophysiological characteristics. For instance, $Orangutan$ represents spike frequency with a fixed numeric value and is insensitive to specific spike patterns. Furthermore, neural spiking activity is not the core foundation for the learning mechanisms planned for subsequent implementation. The main reason is that $Orangutan$ strives to balance biological plausibility with algorithmic validity; its simulation of brain-like mechanisms is constrained by neuroscience on the one hand and driven by algorithmic needs in practical engineering scenarios on the other. To date, there has been no necessity encountered in the development process to implement mechanisms of similar complexity. Should future research present computational problems requiring such mechanisms, I will consider incorporating them into $Orangutan$. Currently, $Orangutan$ is more focused on simulating the chemical properties of neurons, including the regulatory roles of various neurotransmitters in the signal transmission process. Additionally, forthcoming research on learning mechanisms will also be based more on chemical features.
	
	Moreover, $Orangutan$ shares many similarities with other theoretical and computational frameworks within the field of brain-inspired artificial intelligence. For example, the Hierarchical Temporal Memory (HTM) model simulates the brain's cortical column structure at the mesoscopic scale, while the HMAX model emulates the process of hierarchical feature extraction in the brain's ventral pathway structure at the macroscopic level \cite{Riesenhuber1999HierarchicalMO}. $Orangutan$ is dedicated to researching and simulating the process of brain fusion, incorporating a multitude of computing mechanisms. Therefore, beyond the aforementioned mechanisms, Orangutan has implemented additional brain-like mechanisms and has attempted to realize complex brain-like functions based on them, achieving preliminary results.
	
	\section{Conclusion}
	
	In this paper, I proposed an artificial intelligence framework highly inspired and constrained by biology, called $Orangutan$, to explore a viable pathway towards general artificial intelligence. It models various classical mechanisms within the biological brain at multiple scales, exhibiting a high degree of biological plausibility. Subsequently, based on $Orangutan$, I constructed a sensory-motor model. It simulates the process of human eye saccades when observing objects and models the working mechanisms of the classical ventral and dorsal pathways in the biological visual system. The test results of the model on the MNIST dataset demonstrate $Orangutan$'s algorithmic efficacy. This research points to a potential and feasible path for achieving general artificial intelligence close to the human level through multi-scale brain emulation and lays the groundwork for further exploration of artificial intelligence with embodied cognition and sustained learning capabilities.
	
	\section*{Acknowledgements}
	This study utilized OpenAI's ChatGPT tool during the translation process to assist with English translation and text proofreading. I am grateful for the support it provided.
	
	\bibliographystyle{ieeetr}
	\bibliography{example}  
	
	\appendix
	
	\section{Supplementary Information}
	
	\subsection{Saccadic Data on the Test Set}
	\label{sec:saccade_test_result}
	
	Presented here are the complete saccadic trajectory data of the feature observation model on 100 MNIST samples, displayed in the form of key-value pairs: the key is the BMP image filename, and the value is a list of the first five features observed by the model, with their symbols presented in LaTeX format. Each feature's subscript denotes the coordinate of its position.
	
	\begin{lstlisting}[language=Python]
0_0: [
	Arc_{y=15,x=15},
	Ang_{y=21,x=20}^{o1=45,o2=225},
	Ang_{y=8,x=20}^{o1=315,o2=135},
	Ang_{y=23,x=9}^{o1=135,o2=315},
	Ang_{y=7,x=11}^{o1=225,o2=45}
],
0_1: [
	Arc_{y=15,x=15},
	Ang_{y=17,x=7}^{o1=180,o2=360},
	Ang_{y=22,x=21}^{o1=45,o2=225},
	Ang_{y=25,x=13}^{o1=90,o2=270},
	Ang_{y=13,x=24}^{o1=360,o2=180}
],
0_2: [
	Arc_{y=17,x=15},
	Ang_{y=23,x=10}^{o1=135,o2=315},
	Ang_{y=15,x=23}^{o1=360,o2=180},
	Ang_{y=9,x=10}^{o1=225,o2=45},
	Ang_{y=25,x=17}^{o1=90,o2=225}
],
0_3: [
	Arc_{y=14,x=16},
	Ang_{y=20,x=22}^{o1=45,o2=225},
	Ang_{y=8,x=14}^{o1=315,o2=360},
	Ang_{y=9,x=8}^{o1=225,o2=45},
	Ang_{y=23,x=8}^{o1=135,o2=315}
],
0_4: [
	Arc_{y=16,x=15},
	Ang_{y=10,x=12}^{o1=225,o2=315},
	Ang_{y=14,x=22}^{o1=360,o2=180},
	Ang_{y=5,x=14}^{o1=270,o2=90},
	Ang_{y=20,x=8}^{o1=180,o2=360}
],
0_5: [
	Arc_{y=16,x=16},
	Ang_{y=18,x=9}^{o1=180,o2=360},
	Ang_{y=15,x=22}^{o1=360,o2=180},
	Ang_{y=24,x=18}^{o1=45,o2=225},
	Ang_{y=6,x=15}^{o1=270,o2=45}
],
0_6: [
	Arc_{y=14,x=16},
	Ang_{y=10,x=8}^{o1=225,o2=45},
	Ang_{y=9,x=24}^{o1=315,o2=135},
	Ang_{y=20,x=23}^{o1=45,o2=225},
	Ang_{y=21,x=7}^{o1=135,o2=315}
],
0_7: [
	Arc_{y=15,x=15},
	Ang_{y=8,x=9}^{o1=225,o2=45},
	Ang_{y=7,x=22}^{o1=315,o2=135},
	Ang_{y=21,x=23}^{o1=45,o2=225},
	Ang_{y=19,x=7}^{o1=180,o2=337.5}
],
0_8: [
	Arc_{y=17,x=16},
	Ang_{y=9,x=11}^{o1=225,o2=45},
	Ang_{y=22,x=21}^{o1=45,o2=225},
	Ang_{y=8,x=22}^{o1=315,o2=135},
	Ang_{y=22,x=9}^{o1=135,o2=315}
],
0_9: [
	Arc_{y=15,x=14},
	Ang_{y=9,x=20}^{o1=315,o2=135},
	Ang_{y=24,x=15}^{o1=90,o2=270},
	Ang_{y=9,x=7}^{o1=225,o2=45},
	Ang_{y=20,x=24}^{o1=45,o2=180}
],
1_0: [
	Ang_{y=15,x=16}^{o1=22.5,o2=180},
	Ang_{y=19,x=13}^{o1=225,o2=360},
	Ang_{y=9,x=16}^{o1=225,o2=360},
	Ang_{y=24,x=14}^{o1=45,o2=202.5},
	Ang_{y=4,x=18}^{o1=270,o2=90}
],
1_1: [
	Ang_{y=16,x=15}^{o1=45,o2=180},
	Ang_{y=17,x=13}^{o1=225,o2=360},
	Ang_{y=8,x=16}^{o1=225,o2=360},
	Ang_{y=25,x=12}^{o1=135,o2=315},
	Ang_{y=8,x=19}^{o1=45,o2=180}
],
1_2: [
	Ang_{y=11,x=13}^{o1=180,o2=360},
	Ang_{y=16,x=16}^{o1=45,o2=135},
	Ang_{y=9,x=17}^{o1=45,o2=135},
	Ang_{y=24,x=15}^{o1=45,o2=225},
	Ang_{y=4,x=15}^{o1=270,o2=67.5}
],
1_3: [
	Ang_{y=17,x=17}^{o1=360,o2=180},
	Ang_{y=9,x=13}^{o1=202.5,o2=315},
	Ang_{y=17,x=14}^{o1=202.5,o2=315},
	Ang_{y=5,x=14}^{o1=315,o2=135},
	Ang_{y=24,x=14}^{o1=135,o2=315}
],
1_4: [
	Ang_{y=17,x=15}^{o1=180,o2=337.5},
	Ang_{y=13,x=16}^{o1=360,o2=157.5},
	Ang_{y=24,x=18}^{o1=45,o2=135},
	Ang_{y=6,x=12}^{o1=225,o2=360},
	Ang_{y=26,x=17}^{o1=90,o2=270}
],
1_5: [
	Ang_{y=17,x=17}^{o1=360,o2=180},
	Ang_{y=17,x=13}^{o1=180,o2=360},
	Ang_{y=5,x=14}^{o1=270,o2=90},
	Ang_{y=26,x=15}^{o1=135,o2=270}
],
1_6: [
	Ang_{y=17,x=12}^{o1=180,o2=360},
	Ang_{y=12,x=17}^{o1=45,o2=180},
	Ang_{y=6,x=17}^{o1=360,o2=90},
	Ang_{y=24,x=16}^{o1=45,o2=225},
	Ang_{y=5,x=14}^{o1=225,o2=45}
],
1_7: [
	Ang_{y=15,x=16}^{o1=360,o2=180},
	Ang_{y=14,x=14}^{o1=180,o2=360},
	Ang_{y=25,x=14}^{o1=90,o2=270}
],
1_8: [
	Ang_{y=14,x=14}^{o1=180,o2=315},
	Ang_{y=17,x=18}^{o1=360,o2=135},
	Ang_{y=5,x=14}^{o1=315,o2=135},
	Ang_{y=24,x=20}^{o1=45,o2=225},
	Ang_{y=8,x=11}^{o1=180,o2=315}
],
1_9: [
	Ang_{y=19,x=15}^{o1=22.5,o2=180},
	Ang_{y=12,x=15}^{o1=202.5,o2=360},
	Ang_{y=7,x=19}^{o1=360,o2=180},
	Ang_{y=25,x=12}^{o1=135,o2=315},
	Ang_{y=5,x=17}^{o1=270,o2=90}
],
2_0: [
	Ang_{y=20,x=11}^{o1=45,o2=90},
	Ang_{y=24,x=16}^{o1=90,o2=270},
	Arc_{y=8,x=13},
	Ang_{y=5,x=9}^{o1=225,o2=45},
	Ang_{y=8,x=18}^{o1=360,o2=180}
],
2_1: [
	Arc_{y=12,x=15},
	Ang_{y=21,x=15}^{o1=135,o2=225},
	Ang_{y=7,x=9}^{o1=225,o2=45},
	Arc_{y=18,x=21},
	Ang_{y=4,x=17}^{o1=270,o2=90}
],
2_2: [
	Ang_{y=22,x=11}^{o1=45,o2=67.5},
	Arc_{y=12,x=12},
	Ang_{y=22,x=17}^{o1=67.5,o2=225},
	Ang_{y=6,x=12}^{o1=247.5,o2=45},
	Ang_{y=10,x=17}^{o1=45,o2=180}
],
2_3: [
	Ang_{y=16,x=14}^{o1=45,o2=90},
	Ang_{y=19,x=12}^{o1=90,o2=225},
	Ang_{y=9,x=17}^{o1=225,o2=315},
	Ang_{y=18,x=9}^{o1=270,o2=45},
	Ang_{y=5,x=18}^{o1=315,o2=135}
],
2_4: [
	Ang_{y=20,x=16}^{o1=135,o2=225},
	Arc_{y=9,x=14},
	Arc_{y=19,x=12},
	Ang_{y=4,x=17}^{o1=315,o2=135},
	Ang_{y=17,x=16}^{o1=45,o2=112.5}
],
2_5: [
	Arc_{y=13,x=15},
	Ang_{y=23,x=16}^{o1=135,o2=225},
	Ang_{y=21,x=19}^{o1=45,o2=135},
	Ang_{y=4,x=13}^{o1=270,o2=90},
	Arc_{y=21,x=12}
],
2_6: [
	Ang_{y=20,x=15}^{o1=45,o2=90},
	Arc_{y=13,x=13},
	Ang_{y=8,x=13}^{o1=270,o2=90},
	Ang_{y=21,x=14}^{o1=112.5,o2=225},
	Ang_{y=11,x=6}^{o1=225,o2=360}
],
2_7: [
	Arc_{y=12,x=14},
	Ang_{y=20,x=17}^{o1=135,o2=225},
	Ang_{y=16,x=19}^{o1=45,o2=135},
	Ang_{y=5,x=13}^{o1=270,o2=45},
	Ang_{y=8,x=21}^{o1=360,o2=180}
],
2_8: [
	Arc_{y=12,x=15},
	Arc_{y=20,x=12},
	Ang_{y=20,x=18}^{o1=135,o2=225},
	Ang_{y=16,x=22}^{o1=22.5,o2=180},
	Ang_{y=5,x=16}^{o1=270,o2=90}
],
2_9: [
	Arc_{y=13,x=15},
	Ang_{y=20,x=18}^{o1=180,o2=225},
	Ang_{y=18,x=19}^{o1=45,o2=135},
	Ang_{y=9,x=9}^{o1=225,o2=45},
	Ang_{y=21,x=10}^{o1=180,o2=360}
],
3_0: [
	Arc_{y=18,x=16},
	Arc_{y=10,x=10},
	Ang_{y=12,x=20}^{o1=315,o2=135},
	Ang_{y=22,x=11}^{o1=135,o2=315},
	Ang_{y=10,x=5}^{o1=180,o2=360}
],
3_1: [
	Ang_{y=22,x=18}^{o1=45,o2=225},
	Ang_{y=12,x=17}^{o1=45,o2=135},
	Arc_{y=19,x=13},
	Ang_{y=6,x=13}^{o1=270,o2=90},
	Ang_{y=10,x=14}^{o1=247.5,o2=270}
],
3_2: [
	Arc_{y=19,x=15},
	Ang_{y=12,x=16}^{o1=45,o2=135},
	Arc_{y=10,x=12},
	Ang_{y=5,x=12}^{o1=270,o2=90},
	Ang_{y=26,x=13}^{o1=90,o2=270}
],
3_3: [
	Ang_{y=12,x=16}^{o1=360,o2=135},
	Ang_{y=7,x=12}^{o1=180,o2=225},
	Ang_{y=23,x=16}^{o1=45,o2=225},
	Arc_{y=18,x=14},
	Ang_{y=5,x=10}^{o1=225,o2=45}
],
3_4: [
	Arc_{y=18,x=17},
	Ang_{y=6,x=12}^{o1=270,o2=90},
	Ang_{y=11,x=10}^{o1=225,o2=270},
	Ang_{y=24,x=16}^{o1=90,o2=270},
	Ang_{y=12,x=19}^{o1=45,o2=112.5}
],
3_5: [
	Arc_{y=19,x=11},
	Arc_{y=11,x=14},
	Ang_{y=16,x=18}^{o1=45,o2=180},
	Ang_{y=6,x=18}^{o1=315,o2=135},
	Ang_{y=23,x=12}^{o1=90,o2=247.5}
],
3_6: [
	Ang_{y=10,x=16}^{o1=247.5,o2=270},
	Arc_{y=19,x=14},
	Ang_{y=14,x=18}^{o1=45,o2=135},
	Ang_{y=23,x=19}^{o1=45,o2=225},
	Ang_{y=5,x=14}^{o1=270,o2=90}
],
3_7: [
	Arc_{y=19,x=15},
	Arc_{y=9,x=14},
	Ang_{y=25,x=13}^{o1=90,o2=270},
	Ang_{y=13,x=16}^{o1=45,o2=112.5},
	Ang_{y=21,x=20}^{o1=45,o2=180}
],
3_8: [
	Arc_{y=22,x=16},
	Ang_{y=7,x=14}^{o1=225,o2=247.5},
	Arc_{y=17,x=15},
	Ang_{y=22,x=20}^{o1=45,o2=225},
	Ang_{y=12,x=15}^{o1=45,o2=90}
],
3_9: [
	Arc_{y=18,x=15},
	Ang_{y=12,x=19}^{o1=45,o2=135},
	Arc_{y=10,x=15},
	Ang_{y=23,x=19}^{o1=45,o2=225},
	Ang_{y=6,x=14}^{o1=247.5,o2=45}
],
4_0: [
	Arc_{y=13,x=14},
	Ang_{y=19,x=16}^{o1=180,o2=270},
	Ang_{y=19,x=20}^{o1=22.5,o2=180},
	Ang_{y=9,x=10}^{o1=225,o2=360},
	Ang_{y=15,x=7}^{o1=180,o2=360}
],
4_1: [
	Ang_{y=16,x=16}^{o1=225,o2=270},
	Arc_{y=10,x=15},
	Ang_{y=10,x=22}^{o1=45,o2=202.5},
	Ang_{y=7,x=9}^{o1=225,o2=45},
	Ang_{y=23,x=16}^{o1=22.5,o2=45}
],
4_2: [
	Arc_{y=12,x=15},
	Ang_{y=16,x=16}^{o1=180,o2=247.5},
	Ang_{y=14,x=19}^{o1=22.5,o2=135},
	Ang_{y=20,x=18}^{o1=45,o2=180},
	Ang_{y=13,x=9}^{o1=202.5,o2=360}
],
4_3: [
	Arc_{y=13,x=14},
	Ang_{y=18,x=17}^{o1=180,o2=270},
	Ang_{y=18,x=18}^{o1=90,o2=157.5},
	Ang_{y=16,x=19}^{o1=360,o2=90},
	Ang_{y=12,x=9}^{o1=225,o2=360}
],
4_4: [
	Arc_{y=12,x=15},
	Ang_{y=18,x=16}^{o1=180,o2=270},
	Ang_{y=17,x=20}^{o1=22.5,o2=180},
	Ang_{y=12,x=9}^{o1=225,o2=22.5},
	Ang_{y=7,x=14}^{o1=360,o2=135}
],
4_5: [
	Ang_{y=17,x=9}^{o1=45,o2=90},
	Ang_{y=12,x=10}^{o1=225,o2=360},
	Ang_{y=19,x=17}^{o1=135,o2=247.5},
	Ang_{y=12,x=24}^{o1=45,o2=180},
	Ang_{y=11,x=22}^{o1=225,o2=360}
],
4_6: [
	Ang_{y=18,x=18}^{o1=22.5,o2=180},
	Arc_{y=11,x=14},
	Ang_{y=16,x=16}^{o1=202.5,o2=225},
	Ang_{y=17,x=9}^{o1=135,o2=315},
	Ang_{y=11,x=8}^{o1=225,o2=360}
],
4_7: [
	Ang_{y=18,x=17}^{o1=225,o2=270},
	Arc_{y=11,x=15},
	Ang_{y=19,x=20}^{o1=22.5,o2=180},
	Ang_{y=8,x=9}^{o1=225,o2=45},
	Ang_{y=26,x=17}^{o1=90,o2=225}
],
4_8: [
	Ang_{y=19,x=17}^{o1=180,o2=270},
	Arc_{y=12,x=14},
	Ang_{y=17,x=21}^{o1=45,o2=180},
	Ang_{y=11,x=7}^{o1=180,o2=360},
	Ang_{y=13,x=20}^{o1=22.5,o2=45}
],
4_9: [
	Arc_{y=11,x=16},
	Ang_{y=17,x=18}^{o1=225,o2=247.5},
	Ang_{y=14,x=21}^{o1=45,o2=180},
	Ang_{y=12,x=8}^{o1=225,o2=360},
	Ang_{y=24,x=16}^{o1=180,o2=315}
],
5_0: [
	Arc_{y=20,x=16},
	Arc_{y=13,x=15},
	Ang_{y=10,x=8}^{o1=225,o2=45},
	Ang_{y=19,x=11}^{o1=157.5,o2=270},
	Ang_{y=23,x=19}^{o1=45,o2=225}
],
5_1: [
	Arc_{y=18,x=16},
	Ang_{y=10,x=10}^{o1=225,o2=315},
	Arc_{y=9,x=15},
	Ang_{y=25,x=14}^{o1=90,o2=270},
	Ang_{y=15,x=23}^{o1=315,o2=135}
],
5_2: [
	Arc_{y=11,x=18},
	Arc_{y=19,x=14},
	Ang_{y=5,x=21}^{o1=270,o2=90},
	Ang_{y=24,x=13}^{o1=67.5,o2=225},
	Ang_{y=21,x=10}^{o1=225,o2=360}
],
5_3: [
	Arc_{y=19,x=14},
	Ang_{y=12,x=12}^{o1=225,o2=360},
	Arc_{y=10,x=16},
	Ang_{y=23,x=18}^{o1=45,o2=225},
	Ang_{y=6,x=18}^{o1=292.5,o2=90}
],
5_4: [
	Ang_{y=14,x=11}^{o1=67.5,o2=90},
	Arc_{y=19,x=14},
	Ang_{y=8,x=18}^{o1=270,o2=67.5},
	Ang_{y=25,x=11}^{o1=90,o2=270},
	Ang_{y=15,x=8}^{o1=225,o2=360}
],
5_5: [
	Arc_{y=18,x=16},
	Arc_{y=10,x=14},
	Ang_{y=6,x=13}^{o1=270,o2=45},
	Ang_{y=25,x=14}^{o1=90,o2=270},
	Ang_{y=18,x=23}^{o1=360,o2=180}
],
5_6: [
	Ang_{y=10,x=16}^{o1=270,o2=45},
	Ang_{y=20,x=12}^{o1=202.5,o2=315},
	Arc_{y=14,x=15},
	Ang_{y=21,x=14}^{o1=45,o2=180},
	Ang_{y=9,x=23}^{o1=90,o2=225}
],
5_7: [
	Arc_{y=18,x=12},
	Ang_{y=10,x=14}^{o1=90,o2=135},
	Ang_{y=23,x=19}^{o1=45,o2=225},
	Ang_{y=6,x=21}^{o1=270,o2=90},
	Ang_{y=9,x=7}^{o1=225,o2=45}
],
5_8: [
	Arc_{y=19,x=13},
	Arc_{y=11,x=17},
	Ang_{y=6,x=13}^{o1=270,o2=45},
	Ang_{y=25,x=16}^{o1=90,o2=225},
	Ang_{y=12,x=7}^{o1=180,o2=315}
],
5_9: [
	Ang_{y=11,x=12}^{o1=90,o2=135},
	Arc_{y=20,x=13},
	Ang_{y=8,x=15}^{o1=270,o2=45},
	Ang_{y=5,x=22}^{o1=270,o2=90},
	Ang_{y=25,x=11}^{o1=135,o2=270}
],
6_0: [
	Arc_{y=16,x=19},
	Ang_{y=17,x=10}^{o1=22.5,o2=45},
	Arc_{y=20,x=13},
	Ang_{y=11,x=9}^{o1=202.5,o2=360},
	Ang_{y=21,x=21}^{o1=45,o2=225}
],
6_1: [
	Arc_{y=18,x=15},
	Ang_{y=13,x=12}^{o1=45,o2=90},
	Ang_{y=22,x=12}^{o1=135,o2=315},
	Ang_{y=8,x=11}^{o1=225,o2=22.5},
	Ang_{y=22,x=20}^{o1=45,o2=225}
],
6_2: [
	Arc_{y=16,x=18},
	Arc_{y=12,x=12},
	Ang_{y=20,x=10}^{o1=135,o2=315},
	Ang_{y=7,x=9}^{o1=225,o2=360},
	Ang_{y=20,x=21}^{o1=45,o2=225}
],
6_3: [
	Arc_{y=19,x=15},
	Ang_{y=10,x=13}^{o1=225,o2=22.5},
	Ang_{y=14,x=14}^{o1=45,o2=90},
	Ang_{y=22,x=19}^{o1=45,o2=225},
	Ang_{y=16,x=10}^{o1=225,o2=360}
],
6_4: [
	Arc_{y=16,x=20},
	Arc_{y=13,x=12},
	Ang_{y=13,x=6}^{o1=180,o2=360},
	Ang_{y=23,x=16}^{o1=90,o2=270},
	Ang_{y=11,x=20}^{o1=270,o2=90}
],
6_5: [
	Arc_{y=14,x=12},
	Arc_{y=16,x=18},
	Ang_{y=9,x=9}^{o1=202.5,o2=22.5},
	Ang_{y=23,x=14}^{o1=90,o2=270},
	Ang_{y=15,x=22}^{o1=45,o2=180}
],
6_6: [
	Arc_{y=14,x=14},
	Arc_{y=19,x=18},
	Ang_{y=7,x=12}^{o1=225,o2=45},
	Ang_{y=22,x=11}^{o1=135,o2=315},
	Ang_{y=22,x=20}^{o1=45,o2=225}
],
6_7: [
	Ang_{y=19,x=12}^{o1=360,o2=45},
	Arc_{y=18,x=19},
	Ang_{y=12,x=9}^{o1=180,o2=360},
	Ang_{y=21,x=22}^{o1=45,o2=225},
	Ang_{y=13,x=23}^{o1=315,o2=135}
],
6_8: [
	Arc_{y=18,x=17},
	Ang_{y=5,x=14}^{o1=225,o2=45},
	Arc_{y=11,x=16},
	Ang_{y=20,x=11}^{o1=135,o2=315},
	Ang_{y=18,x=21}^{o1=45,o2=180}
],
6_9: [
	Arc_{y=13,x=16},
	Ang_{y=8,x=12}^{o1=225,o2=45},
	Arc_{y=21,x=15},
	Ang_{y=22,x=10}^{o1=135,o2=315},
	Ang_{y=4,x=21}^{o1=315,o2=135}
],
7_0: [
	Ang_{y=11,x=20}^{o1=225,o2=270},
	Ang_{y=20,x=17}^{o1=45,o2=202.5},
	Ang_{y=8,x=17}^{o1=270,o2=90},
	Ang_{y=23,x=12}^{o1=225,o2=22.5},
	Ang_{y=10,x=8}^{o1=135,o2=270}
],
7_1: [
	Ang_{y=12,x=20}^{o1=225,o2=270},
	Ang_{y=22,x=18}^{o1=45,o2=225},
	Ang_{y=7,x=10}^{o1=270,o2=90},
	Ang_{y=23,x=13}^{o1=225,o2=360},
	Ang_{y=10,x=8}^{o1=135,o2=225}
],
7_2: [
	Ang_{y=10,x=17}^{o1=180,o2=247.5},
	Ang_{y=16,x=20}^{o1=360,o2=180},
	Ang_{y=24,x=16}^{o1=225,o2=360},
	Ang_{y=8,x=15}^{o1=270,o2=90},
	Ang_{y=11,x=5}^{o1=225,o2=45}
],
7_3: [
	Ang_{y=10,x=19}^{o1=225,o2=270},
	Ang_{y=23,x=15}^{o1=45,o2=225},
	Ang_{y=8,x=12}^{o1=315,o2=67.5},
	Ang_{y=20,x=10}^{o1=225,o2=45},
	Ang_{y=11,x=22}^{o1=45,o2=180}
],
7_4: [
	Ang_{y=18,x=18}^{o1=45,o2=90},
	Arc_{y=14,x=13},
	Ang_{y=21,x=17}^{o1=90,o2=180},
	Ang_{y=7,x=14}^{o1=270,o2=90},
	Ang_{y=24,x=14}^{o1=180,o2=315}
],
7_5: [
	Arc_{y=13,x=15},
	Ang_{y=17,x=17}^{o1=45,o2=90},
	Ang_{y=21,x=16}^{o1=90,o2=202.5},
	Ang_{y=8,x=13}^{o1=270,o2=67.5},
	Ang_{y=20,x=15}^{o1=225,o2=315}
],
7_6: [
	Ang_{y=8,x=11}^{o1=270,o2=90},
	Arc_{y=17,x=11},
	Ang_{y=22,x=20}^{o1=45,o2=180},
	Ang_{y=14,x=21}^{o1=360,o2=135},
	Ang_{y=28,x=16}^{o1=90,o2=270}
],
7_7: [
	Arc_{y=12,x=16},
	Ang_{y=19,x=16}^{o1=90,o2=202.5},
	Ang_{y=15,x=20}^{o1=45,o2=90},
	Ang_{y=7,x=11}^{o1=270,o2=90},
	Ang_{y=17,x=13}^{o1=225,o2=270}
],
7_8: [
	Ang_{y=12,x=20}^{o1=225,o2=270},
	Ang_{y=8,x=10}^{o1=270,o2=90},
	Ang_{y=23,x=17}^{o1=45,o2=225},
	Ang_{y=11,x=5}^{o1=90,o2=270},
	Ang_{y=24,x=11}^{o1=225,o2=45}
],
7_9: [
	Ang_{y=19,x=15}^{o1=202.5,o2=270},
	Arc_{y=14,x=13},
	Ang_{y=15,x=17}^{o1=45,o2=180},
	Ang_{y=9,x=16}^{o1=270,o2=45},
	Ang_{y=18,x=20}^{o1=90,o2=225}
],
8_0: [
	Arc_{y=12,x=16},
	Ang_{y=17,x=13}^{o1=270,o2=315},
	Arc_{y=19,x=12},
	Ang_{y=16,x=18}^{o1=67.5,o2=180},
	Ang_{y=6,x=18}^{o1=270,o2=90}
],
8_1: [
	Arc_{y=21,x=16},
	Ang_{y=15,x=17}^{o1=45,o2=135},
	Arc_{y=11,x=15},
	Ang_{y=16,x=11}^{o1=180,o2=270},
	Ang_{y=8,x=13}^{o1=270,o2=45}
],
8_2: [
	Arc_{y=20,x=13},
	Arc_{y=9,x=16},
	Ang_{y=15,x=11}^{o1=225,o2=360},
	Ang_{y=14,x=17}^{o1=45,o2=135},
	Ang_{y=24,x=13}^{o1=90,o2=225}
],
8_3: [
	Arc_{y=21,x=15},
	Ang_{y=17,x=11}^{o1=225,o2=315},
	Ang_{y=6,x=16}^{o1=270,o2=90},
	Ang_{y=15,x=17}^{o1=45,o2=135},
	Arc_{y=11,x=15}
],
8_4: [
	Ang_{y=15,x=13}^{o1=225,o2=315},
	Arc_{y=21,x=14},
	Ang_{y=14,x=19}^{o1=45,o2=180},
	Arc_{y=10,x=17},
	Ang_{y=7,x=14}^{o1=270,o2=45}
],
8_5: [
	Arc_{y=12,x=14},
	Ang_{y=16,x=15}^{o1=67.5,o2=135},
	Arc_{y=21,x=14},
	Ang_{y=17,x=10}^{o1=180,o2=315},
	Ang_{y=8,x=11}^{o1=225,o2=45}
],
8_6: [
	Arc_{y=20,x=16},
	Arc_{y=10,x=16},
	Ang_{y=14,x=19}^{o1=45,o2=135},
	Ang_{y=17,x=11}^{o1=180,o2=315},
	Ang_{y=24,x=18}^{o1=45,o2=225}
],
8_7: [
	Ang_{y=15,x=13}^{o1=225,o2=315},
	Ang_{y=15,x=17}^{o1=45,o2=135},
	Arc_{y=20,x=15},
	Arc_{y=10,x=15},
	Ang_{y=25,x=15}^{o1=90,o2=270}
],
8_8: [
	Arc_{y=19,x=16},
	Arc_{y=10,x=16},
	Ang_{y=15,x=18}^{o1=45,o2=135},
	Ang_{y=14,x=12}^{o1=225,o2=315},
	Ang_{y=5,x=14}^{o1=270,o2=67.5}
],
8_9: [
	Arc_{y=19,x=15},
	Arc_{y=9,x=15},
	Ang_{y=13,x=19}^{o1=45,o2=135},
	Ang_{y=13,x=14}^{o1=225,o2=270},
	Ang_{y=22,x=18}^{o1=45,o2=225}
],
9_0: [
	Ang_{y=16,x=16}^{o1=157.5,o2=225},
	Arc_{y=13,x=13},
	Ang_{y=17,x=19}^{o1=360,o2=135},
	Ang_{y=8,x=11}^{o1=225,o2=45},
	Ang_{y=9,x=15}^{o1=360,o2=90}
],
9_1: [
	Ang_{y=17,x=15}^{o1=225,o2=270},
	Arc_{y=12,x=16},
	Ang_{y=22,x=16}^{o1=45,o2=225},
	Arc_{y=12,x=20},
	Ang_{y=8,x=17}^{o1=270,o2=90}
],
9_2: [
	Arc_{y=13,x=15},
	Ang_{y=17,x=16}^{o1=202.5,o2=225},
	Ang_{y=17,x=18}^{o1=45,o2=180},
	Ang_{y=10,x=11}^{o1=225,o2=45},
	Ang_{y=9,x=19}^{o1=315,o2=45}
],
9_3: [
	Arc_{y=13,x=14},
	Ang_{y=19,x=16}^{o1=202.5,o2=270},
	Ang_{y=19,x=19}^{o1=22.5,o2=180},
	Ang_{y=8,x=10}^{o1=225,o2=45},
	Ang_{y=9,x=19}^{o1=360,o2=45}
],
9_4: [
	Ang_{y=14,x=17}^{o1=225,o2=247.5},
	Arc_{y=12,x=14},
	Ang_{y=17,x=19}^{o1=45,o2=202.5},
	Ang_{y=10,x=11}^{o1=225,o2=45},
	Ang_{y=8,x=20}^{o1=315,o2=90}
],
9_5: [
	Ang_{y=16,x=19}^{o1=225,o2=247.5},
	Arc_{y=13,x=14},
	Ang_{y=20,x=19}^{o1=45,o2=180},
	Ang_{y=11,x=8}^{o1=225,o2=45},
	Ang_{y=11,x=21}^{o1=360,o2=135}
],
9_6: [
	Arc_{y=14,x=14},
	Ang_{y=21,x=18}^{o1=360,o2=180},
	Ang_{y=22,x=15}^{o1=180,o2=315},
	Ang_{y=7,x=17}^{o1=292.5,o2=90},
	Ang_{y=11,x=9}^{o1=180,o2=360}
],
9_7: [
	Ang_{y=16,x=14}^{o1=225,o2=315},
	Ang_{y=23,x=14}^{o1=45,o2=225},
	Arc_{y=11,x=16},
	Ang_{y=6,x=17}^{o1=270,o2=90},
	Ang_{y=13,x=21}^{o1=45,o2=180}
],
9_8: [
	Arc_{y=11,x=17},
	Ang_{y=16,x=14}^{o1=225,o2=270},
	Ang_{y=9,x=13}^{o1=225,o2=45},
	Ang_{y=20,x=16}^{o1=45,o2=202.5},
	Ang_{y=9,x=20}^{o1=360,o2=135}
],
9_9: [
	Arc_{y=12,x=16},
	Ang_{y=19,x=12}^{o1=225,o2=315},
	Ang_{y=21,x=14}^{o1=45,o2=202.5},
	Ang_{y=8,x=14}^{o1=225,o2=45},
	Ang_{y=7,x=21}^{o1=315,o2=45}
]
\end{lstlisting}
	
\end{document}